\documentclass[aps,pre,onecolumn]{revtex4}
\makeatletter
\newcommand\at[2]{\left.#1\right|_{#2}}
\renewcommand\paragraph{\@startsection{paragraph}{4}{\z@}%
            {-2.5ex\@plus -1ex \@minus -.25ex}%
            {1.25ex \@plus .25ex}%
            {\normalfont\normalsize\bfseries}}
\makeatother

\usepackage{algpseudocode}
\usepackage{amsmath}
\usepackage{mathtools}
\usepackage{graphicx}
\usepackage{slashed}
\usepackage{lineno}
\usepackage{latexsym}
\usepackage{subcaption}
\usepackage{amssymb}

\newcommand{\bb}{\begin{equation}}
\newcommand{\ee}{\end{equation}}

\usepackage{multirow}
\usepackage{ctable}
\usepackage{bm}
\usepackage{enumerate}
\newcommand{\D}[2]{\frac{\partial #1}{\partial #2}}
\newcommand{\DD}[2]{\frac{\partial^2 #1}{\partial #2^2}}

\usepackage{framed}

\newcommand{\fig}[1]{Figure \ref{#1}}

\newcommand{\sect}[1]{Section \ref{#1}}

\newcommand{\eqn}[1]{equation \eqref{#1}}
\newcommand{\eqns}[2]{equations \eqref{#1} and \eqref{#2}}
\newcommand{\eqnto}[2]{equations \eqref{#1}-\eqref{#2}}
\usepackage{url}
\usepackage{soul}

\DeclareMathAlphabet{\pazocal}{OMS}{zplm}{m}{n}

\DeclareMathOperator*{\argmax}{argmax} % thin space, limits underneath in displays
\newcommand{\unif}{\pazocal{U}}

\usepackage[geometry]{ifsym}
\makeatletter
\newcommand\restr[2]{{% we make the whole thing an ordinary symbol
  \left.\kern-\nulldelimiterspace % automatically resize the bar with \right
  #1 % the function
  \vphantom{\big|} % pretend it's a little taller at normal size
  \right|_{#2} % this is the delimiter
  }}
\def\url@leostyle{%
  \@ifundefined{selectfont}{\def\UrlFont{\sf}}{\def\UrlFont{\small\ttfamily}}}
\makeatother
\usepackage{multirow}
\usepackage{blkarray}
\usepackage{soul}
\usepackage{framed}
\usepackage{color}
\usepackage{setspace}

\newcommand{\Dd}[3]{\frac{\partial^2 #1}{\partial #2\partial #3}}
\usepackage[]{natbib} 
\graphicspath{ {./plots/} }

\begin{document}
\title{Coupling Bayesian theory and static acoustic detector data to model bat motion and locate roosts}
\author{Lucy Henley\footnote[1]{HenleyL1@cardiff.ac.uk}}
\author{Owen Jones}
\affiliation{Cardiff School of Mathematics
Cardiff University
Senghennydd Road, Cardiff, CF24 4AG, UK}%
\author{Fiona Mathews}
\affiliation{University of Sussex, John Maynard Smith Building, Falmer, Sussex, BN1 9QG, UK}%

\author{Thomas E. Woolley}
\affiliation{Cardiff School of Mathematics
Cardiff University
Senghennydd Road, Cardiff, CF24 4AG, UK}%

\date{\today}
\begin{abstract}
We propose a novel approach for modelling bat motion dynamics and use it to predict roost locations using data from static acoustic detectors. Specifically, radio tracking studies of Greater Horseshoe bats demonstrate that bat movement can be split into two phases: dispersion and return. Dispersion is easily understood and can be modelled as simple random motion. The return phase is much more complex, as it requires intelligent directed motion and results in all agents returning home in a stereotypical manner. Critically, combining reaction-diffusion theory and domain shrinking we deterministically and stochastically model a ``leap-frogging'' motion, which fits favourably with the observed tracking data. 
\end{abstract}

\maketitle

\section{Introduction}
\label{chap:intro}

  Bats play an important role in the UK ecosystem, as they help control insect
  populations \cite{Kunz2011} and act as ecological indicators of biodiversity and pollution \cite{Jones2009}.  However, they are susceptible to human impacts due to their sensitivity to
  light, noise and temperature. Additionally, habitat fragmentation due to roads
  and building work can reduce foraging opportunities and lead to a significant
  risk of population decline \cite{rossiter2000genetic}.
 As a result, bats are protected by law in
Europe under the EUROBATS agreement \cite{Eurobats} and domestic law. Therefore, identifying
roosts and important foraging areas is an important step in ensuring that
habitats remain protected.

We will use mathematics in a number of ways to solve current problems regarding understanding bat movement, and subsequently use the models we derive to help identify the locations of bat roosts. We will consider the movement of Greater Horseshoe Bats, a species that is classified as Near Threatened across Europe owing to significant declines in its distribution and abundance over the last 50 years.  Great Britain, and particularly south-west England are a stronghold for the species, and several detailed ecological studies have been conducted in this region.

  Radio tracking surveys are commonly used to identify the movement and habitat use of bats
   by using radio transmitters to locate bats \cite{Bontadina2002,Encarnacao2005,kunz1988ecological}.
  In order to track bats, they must first be caught using a humane method, such as a harp trap.
 Once a bat is caught, a small radio  transmitter is attached to its back, often using surgical glue, and it is
\begin{figure}
     \centering
     \begin{subfigure}[b]{0.45\textwidth}
         \centering
         \includegraphics[width=\textwidth]{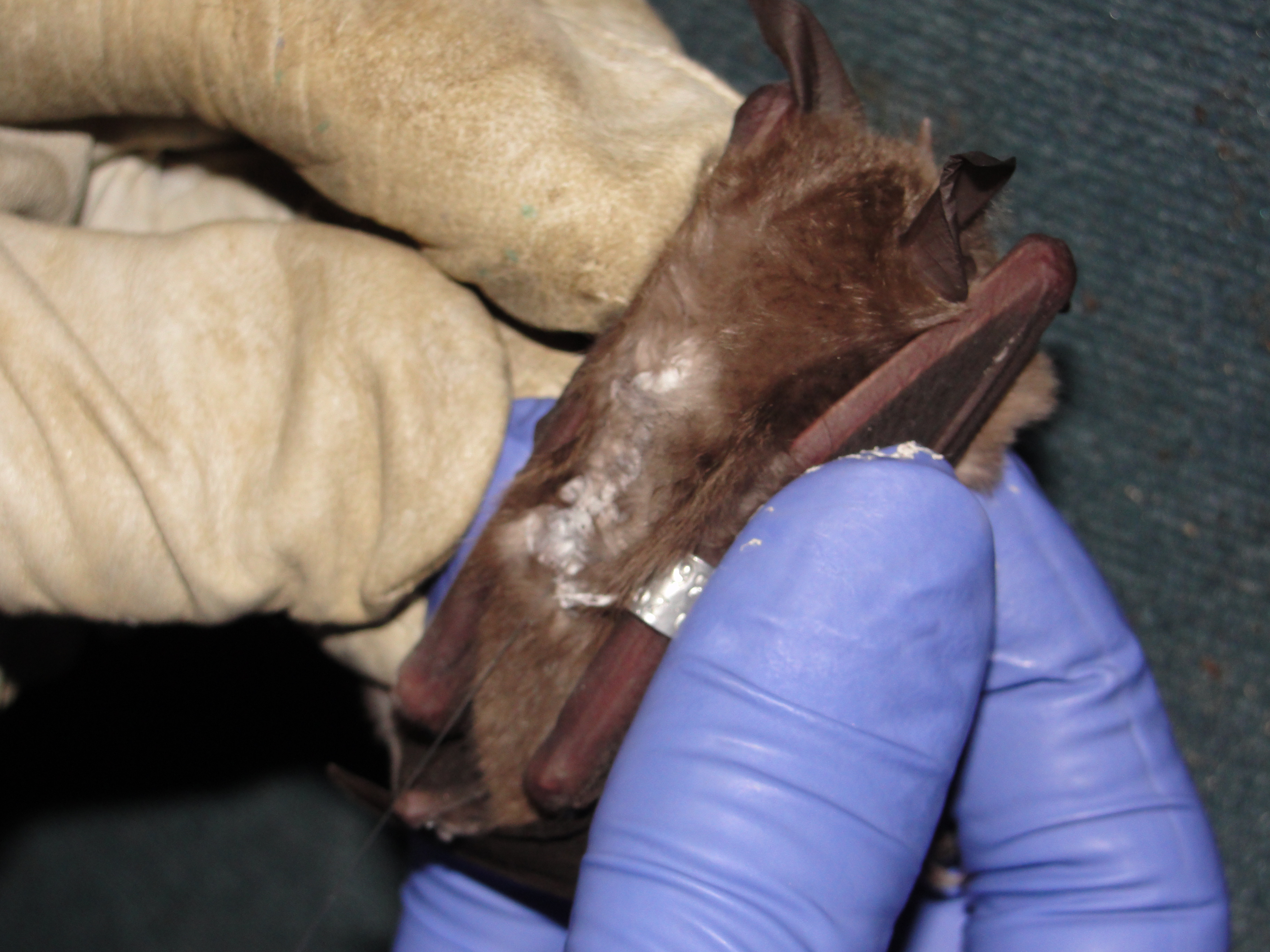}
     \end{subfigure}
     \hfill
     \begin{subfigure}[b]{0.45\textwidth}
         \centering
         \includegraphics[width=\textwidth]{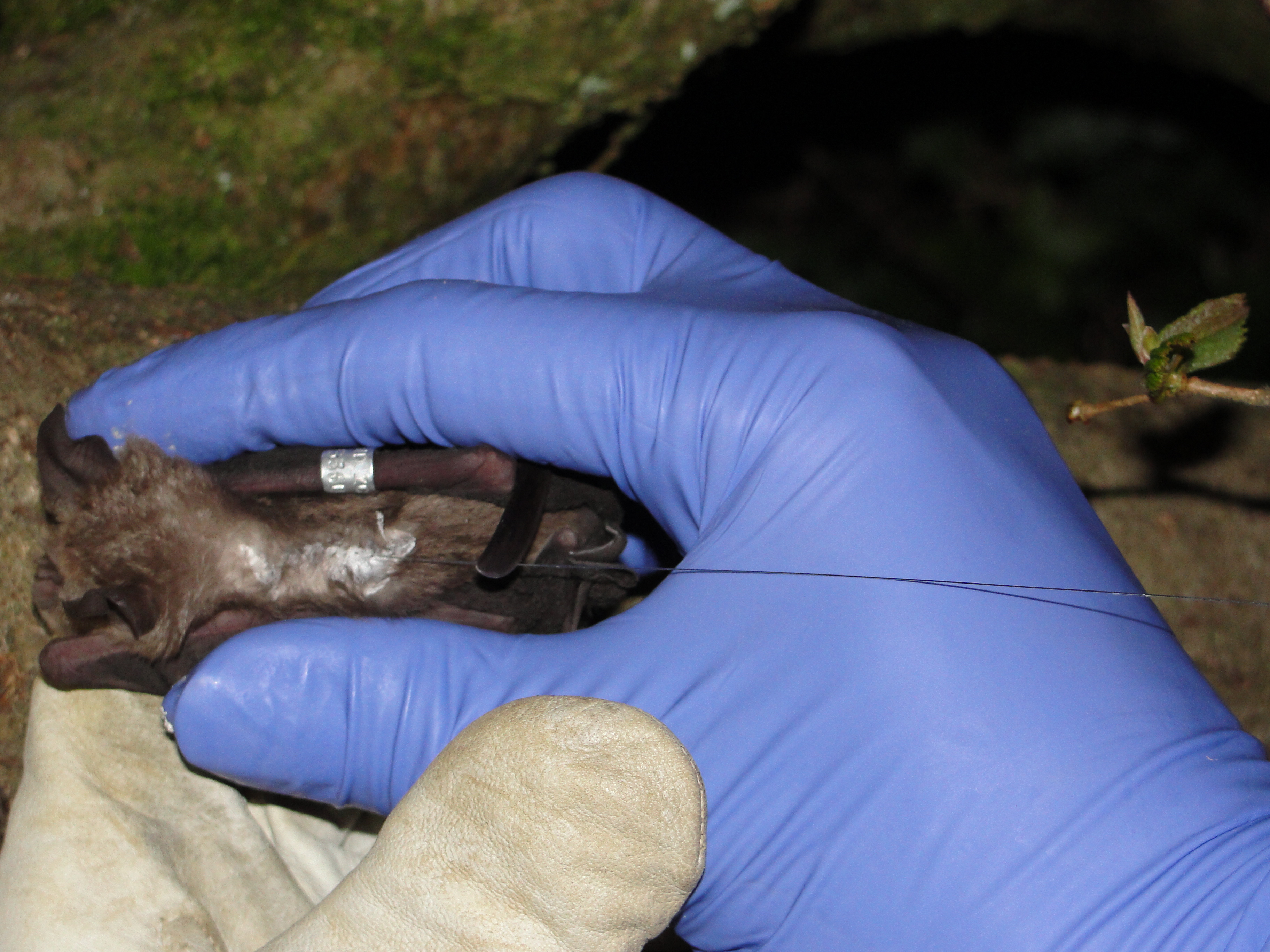}
     \end{subfigure}
     \caption{A Greater Horseshoe bat with radio transmitter glued to its back.}
     \label{fig:bat_tracker}
\end{figure}
 released. The transmitter must be less than 5\% of the bat's weight in order to avoid disrupting flight patterns \cite{brigham1988load}. Images of a typical transmitter attached to a Greater Horseshoe bat are shown in \fig{fig:bat_tracker}. The
  signal from a
transmitter is then picked up by field
 workers using scanning radio-receivers. The precise position of the bat can be difficult to determine: bats can fly at speeds of around 30mph, and it can be difficult to get a bearing sufficiently synchronously to calculate the location accurately. As such, the location of the bat is
 estimated by triangulation. Field workers follow the bat and attempt to maintain contact
 throughout the night, taking regular recordings of location until either the signal is
 lost, or the bat returns to
the roost. Due to the nature of the tracking,
 locations are not recorded at regular intervals, rather only when the signal is
 found. Although useful, radio
tracking surveys are highly labour intensive as
 they require first locating a
roost in order to catch bats, and then teams of
 workers following each bat over the entire night. Transmitters have a limited
 range, and workers therefore
must remain close to the bat in order to pick up
 the signal, which can often be
difficult in a rural environment with obstacles
 such as impassable waterways
and hedgerows. Additionally,
  transmitters have a
 limited battery life and are
often detached and lost before the end of the
 survey, meaning that surveys can
be cut short.

It can be useful to derive mathematical models using the data collected from radio tracking studies in order to gain more insight into bat's behaviour. These models can also be useful in identifying bat roosts. In the maternity season, hundreds of animals can group together in a single roost to have young, and as such it is imperative that these roosts are identified and protected. However, locating roosts is difficult because they tend to be in buildings or underground, so radio-signals of tagged bats tend to lost. Additionally, there are generally numerous potential roost locations, and it is not possible to conduct physical surveys of each site. Acoustic surveys using acoustic detectors are widely used to study bats, however, they have the important disadvantage that although they record the  presence of bats in flight, they cannot record flight direction or used to locate roosts in the same way that radio tracking studies can. Mathematical models can help us to interpret acoustic records and can allow us to gain more insight into the presence of possible roosts, as we would expect many more records close to a roost than at detectors further away.

Mathematical models are an
invaluable tool in understanding
ecological
mechanisms as they help us to
understand the ecological mechanisms
that lead to certain patterns in behaviour
\cite{Ovaskainen2016}. There are
many possible formalisms, such as stochastic,
deterministic, spatial, or
discrete depending on the population and behaviour
\cite{Murray2011}.  Here we will focus on deterministic models, as these can provide a useful approximation to real life whilst simplifying the mathematics by excluding noise.

We combine statistical methods and deterministic modelling to characterise bat motion. We use time-location data from radio tracking studies that track bat motion from when they first leave their roost at sunset to when they return in the morning. By extracting the mean squared displacement from the data we see two distinct movement phases, an initial linear dispersal followed by a gradual return to the roost. We use this data to develop deterministic models to describe motion for each movement phase. Diffusion models in both one and two dimensions are discussed to describe the dispersal of bats away from the roosts. Two models are used to describe movement for the remainder of the night whilst bats are foraging, a convection-diffusion model and a model describing diffusion on a shrinking domain. Convection-diffusion models are widely used in ecology to model population migration, however we will show here that a convection-diffusion model is not consistent with radio tracking data. Instead, a shrinking domain diffusion model provides a better description of bat movement whilst foraging.

\section{Mathematical framework}

In this section we will discuss the general framework of Approximate Bayesian Computation, which will be adapted in multiple ways to: predict bat movement characteristics, fit models of bat movement to radio tracking data and to estimate the location of a roost given bat survey data.

 \subsection{Bayesian statistics}
 Bayesian statistics is a statistical
paradigm that provides useful mathematical tools for updating knowledge about a
 parameter
given related data
\cite{Gelman2013}. Here, we will use Bayesian inference to fit various mathematical movement models to bat survey data and later to fit models to data collected from static detectors to estimate the location of roosts. The probability distribution of a
parameter
 $\theta$
conditioned on observations $\bm{Y} = \{ Y_1, Y_2, ..., Y_n
\}$ is
 given by
Bayes' Theorem,

 \begin{equation}
   p(\theta \mid \bm{Y})
= \frac{p(\bm{Y} \mid \theta) p(\theta)}{p(\bm{Y})}
   \nonumber,
	 \label{eqn:bayes_rule}
 \end{equation}
 where
$p(\theta \mid \bm{Y})$ is the posterior probability distribution, formally
describing the probability that the parameter value is $\theta$ given
observations $\bm{Y}$. The likelihood function describes the probability of observing $\bm{Y}$ if the parameter value is $\theta$ and is given by $p(\bm{Y} \mid \theta)$.
The
 prior distribution is $p(\theta)$, describing the initial knowledge of
possible
 parameter values.

This method of Bayesian inference is extremely useful for problems with a calculable likelihood function. However, for many problems, including those discussed in this paper, the likelihood function is intractable. In this case, a useful approach is to use Approximate Bayesian Computation.

 \subsubsection{Approximate Bayesian Computation (ABC)} \label{sect:abc}
 Approximate Bayesian Computation (ABC) is an approach to Bayesian
inference using
 simulation and random sampling \cite{Beaumont2002,
Sisson2010} and is widely used for problems where the analytical form of the likelihood function is intractable \cite{myung2003tutorial}. We will use ABC to fit mathematical movement models to bat survey data.
 ABC replaces the calculation
 of the likelihood function
$p(\bm{Y}
\mid \theta)$ with simulation of a model using a specific parameter
value
$\theta'$ to produce an artificial dataset
 $\bm{X}$. Then, some
distance metric,
$\rho (\bm{X}, \bm{Y}) $, usually defined
 as a distance
between summary
statistics of $\bm{X}$ and $\bm{Y}$, is used to
 compare
simulated data $\bm{X}$
to observations $\bm{Y}$. If $\rho (\bm{X},
 \bm{Y}) $
is smaller than some
threshold value, $\epsilon$, the simulated data is
 close
enough to observations
that the candidate parameter $\theta'$ has some
nonzero probability of being in
the posterior distribution $ p(\theta \mid \bm{Y})$, and the sample $\theta'$
is accepted into the simulated posterior
distribution. This is repeated until
the desired sample size is reached. For
small $\epsilon$, assuming that the
model is correct, the simulated posterior
distribution produced approximates
the
 true posterior $ p(\theta \mid
\bm{Y})$ \cite{Sisson2010}, and this simulated posterior provides a probability distribution for the parameter $\theta$.

A pseudocode for the ABC method is as
follows. First, the threshold parameter $\epsilon$ and sample size $n$ are fixed and the
 prior distribution for parameters $\theta$ is set, $p(\theta)$.
 \begin{algorithmic}
\While {$i < n$}
     \State Sample $\theta'$ from $p(\theta)$
     \State Simulate $\bm{X}$ from $\theta '$
     \State $\rho \gets \mid \bm{X} - \bm{Y} \mid$
\If {$\rho < \epsilon$}
  \State $\theta_i \gets \theta '$
  \State $i \gets i + 1$
\EndIf
\EndWhile
 \end{algorithmic}
The parameters $\theta_i$ are an approximate sample from the posterior distribution $p(\theta \mid \bm{Y})$, and the posterior mean $\mathbb{E}(\theta \mid \bm{Y})$ can be estimated using

\begin{equation}
\mathbb{E}(\theta \mid \bm{Y}) = \frac{1}{n} \sum_i \theta_i .
\end{equation}

ABC will be used here to fit parameters to the data extracted from radio tracking data. The mean squared displacement from the roost calculated from the radio tracking data will be used as the observations $\bm{Y}$. The mean squared displacement from diffusion simulations will be used as the simulated data $\bm{X}$. The parameters in the shrinking domain diffusion model are the diffusion coefficient, describing the rate of spread, and the shrinking speed of the domain, and these will be used as the parameters $\theta$.

\section{Radio tracking survey} \label{sect:radiotrack}
In this section our aim is to use data gained from a radio tracking survey to produce
a mathematical formulation of bat movement which qualitatively and quantitatively
matches the data. The data suggests that Greater Horseshoe bat movement can be split
into 2 phases during the night. We aim to provide a mechanistic description of these 2
phases, which allow us to understand and describe the motion. Later, we will use this
model of motion to help estimate the most likely locations of Greater Horseshoe bat
roost.

A radio tracking study was conducted at 3 Greater Horseshoe bat roosts in Devon to
study the usage of land surrounding the roosts \cite{Mathews2009}. 12 bats were fitted with
radio tags and studied over 24 nights. Due to a limited number of workers and
 limited battery life on the tags, each bat was not tracked every night. Four day
 roosts were used by bats in the study, with some bats using different roosts on
 different days. The roost used by each bat was not identified on every night.

\begin{figure} [h]
    \centering
        \includegraphics[width=0.8\textwidth]{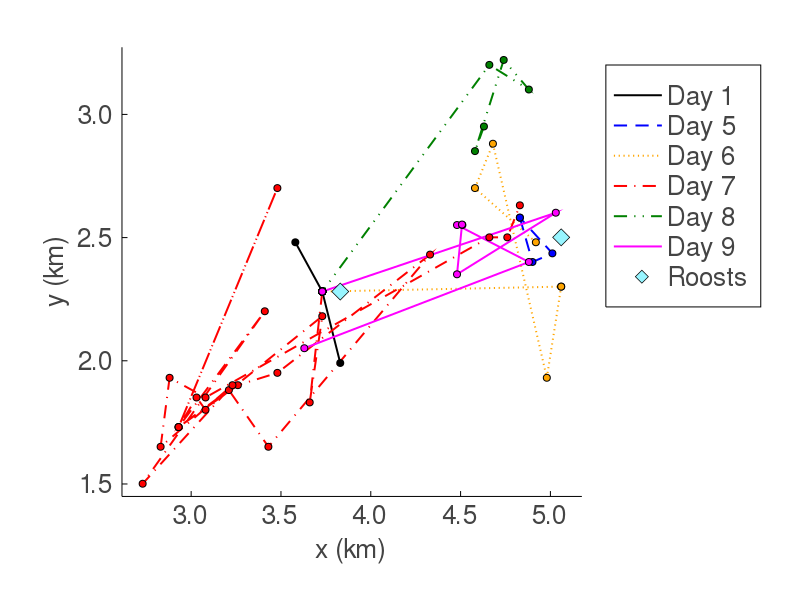}
        \caption{The locations of bat 1 over 6 nights of the survey. The two roosts bat 1 used during the course of the study are shown as diamonds.}
    \label{fig:bat1}
\end{figure}

\begin{figure}
\centering
    \includegraphics[width=0.8\textwidth]{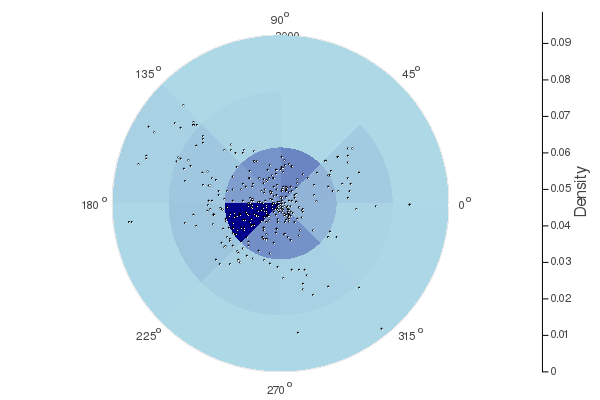}
    \caption{Polar histogram of bat locations over the study period. The density for each segment is calculated as the number of locations recorded in the segment divided by the area of the segment. A scatter plot displaying each recording is overlaid on the histogram.}
\label{fig:polar_histogram}
\end{figure}

 For this analysis, only the
data from nights when a bat's roost was known was used, since the dispersal
from the roost is important.
A total of 322 bat locations were used for this analysis. The trajectory of a bat, referred to here as bat
1, over the 6 nights it was tracked is shown in \fig{fig:bat1}. Bat 1 used 2 of the 4 roosts identified during the course of the study and visited different areas whilst foraging, taking a different route on each night. The number of
locations recorded varies each night because the signal was lost during the night
 on some nights. The recorded locations of all bats over the study period are displayed in a polar histogram in \fig{fig:polar_histogram}. This histogram shows that most records are located close to the roost, and are distributed over all angles.

\begin{figure} [h]
    \centering
        \includegraphics[width=0.6\textwidth]{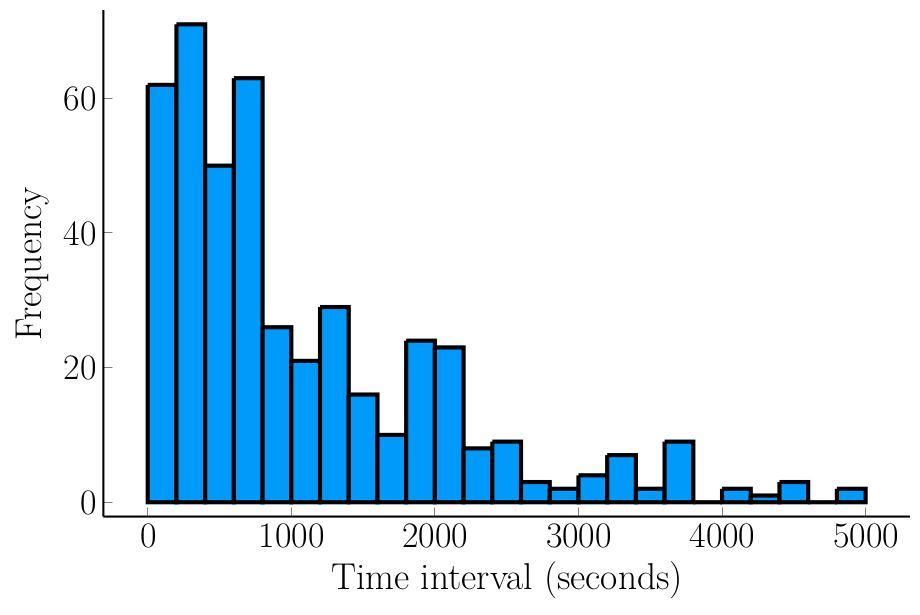}
        \caption{A histogram of the time intervals between consecutive recordings. Outliers with time intervals over the 75th percentile at 5000 seconds have been removed.}
    \label{fig:time_interval}
\end{figure}

\begin{figure} [h]
    \centering
        \includegraphics[width=\textwidth]{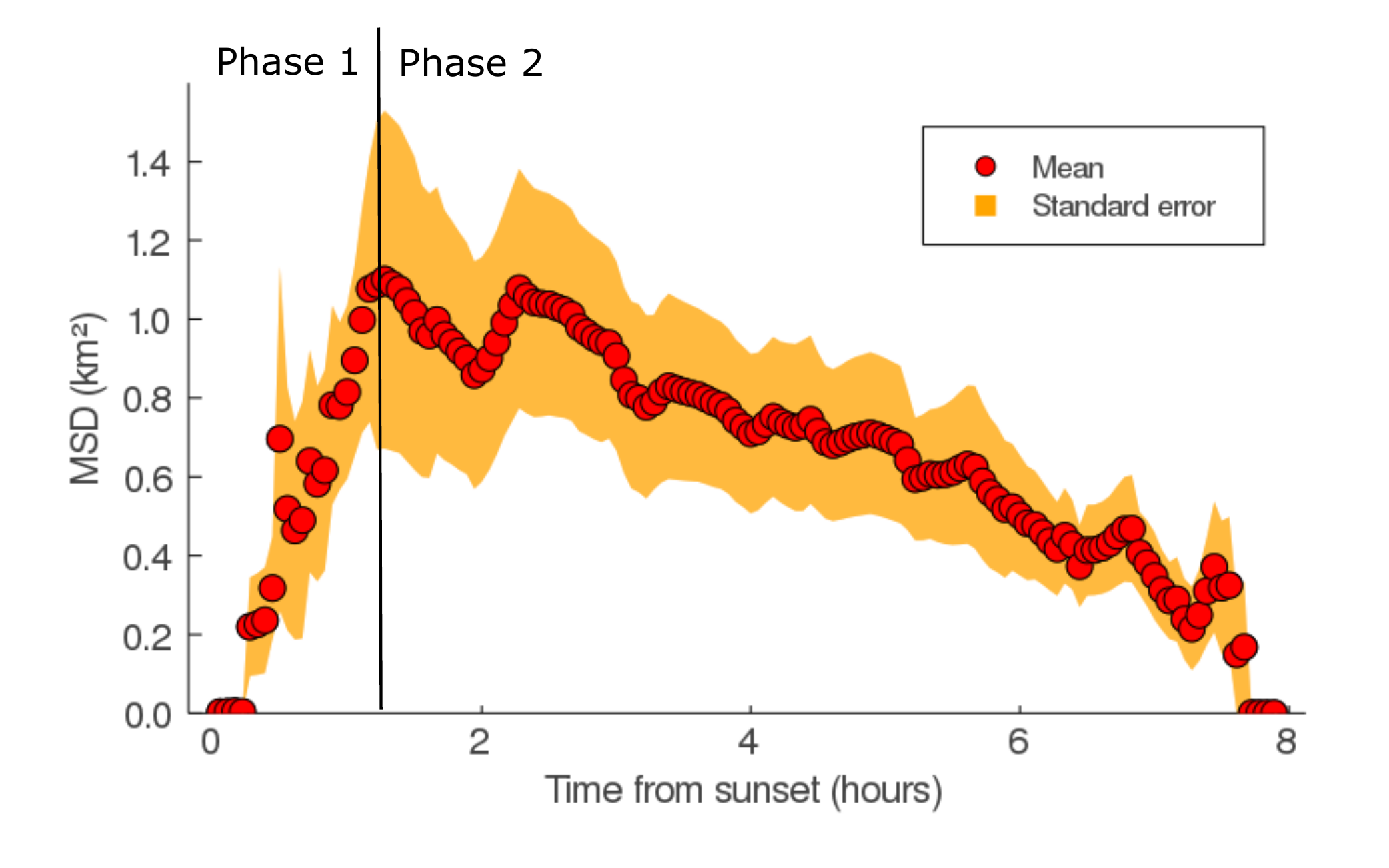}
        \caption{The mean-squared distance (MSD) for all radio tracked bats.
        The red dots are the averaged values over 56 trajectories and the standard error is shown as an orange ribbon.}
    \label{fig:MSD}
\end{figure}

A histogram of time intervals between consecutive recordings is shown in \fig{fig:time_interval}, demonstrating the irregularity of recording intervals. We will use the mean-squared distance (MSD) from the roost as a function of time to summarise the data and this will be used as a measure to fit the movement models to the data. In order to calculate the MSD at a given time, we require regularly spaced recordings. The locations were linearly
interpolated between recordings at intervals of $\Delta t = 200$ seconds, as the distribution in \fig{fig:time_interval} peaks between 100 and 200 seconds.
The MSD from the roost was calculated from the interpolated positions using
\begin{equation}
\left<r^2(t)\right> = \frac{1}{N} \sum_{i=1}{N} |\bm{x_i}(t)-\bm{x_i}(0)|^2,
\end{equation}
where $\bm{x_i}(t)$ is the location $(x,y)$ of bat $i$ at time $t$. The MSD is shown in \fig{fig:MSD}. The standard error is given by

\begin{equation}
\sigma_{\bar{\bm{x}}}  = \frac{\sqrt{\Sigma (\bm{x_i} -  \bm{\bar{x}} )^2}}{n}
\end{equation}

 where $n$ is the number of observations and $\bar{x}$ is the mean location, $\bar{x} = \frac{\Sigma_i{x_i}}{n}$. The standard error is
shown as an orange ribbon around the MSD in \fig{fig:MSD}. On some nights, some bats did not return to the same roost they used the day before and the roost that each bat returned to at the end of the night is not known in the majority of cases, as the signal is lost throughout the night. As a result, we have assumed for the purpose of calculating the MSD that bats have returned to the same roost. As shown in \fig{fig:MSD}, the MSD reaches 0 by the end of the night, suggesting that the recordings from the end of the night are all from bats that returned to the same roost at the end of the night.
The data indicates two movement phases, an initial rapid dispersal from the roosts, followed by a gradual return whilst bats are foraging.

During phase 1, for $0 \leq t < 1.6$ hours, the MSD seems to increase linearly as bats are dispersing. The standard error grows during this phase as the bats spread out. During phase 2, for $1.6 \leq t < 8$ hours, the MSD decreases at an increasing rate as bats move back towards the roost, shrinking to zero at $t \approx 8$ hours. The variation shrinks to zero during this phase as bats start to converge on the roost.
%The standard error grows during this phase because the number of recordings $n$ increases as tagged bats are found and the signal is picked up. During phase 2, for $1.6 \leq t < 8$ hours, the mean squared distance decreases at an increasing rate as bats move back towards the roost, shrinking to zero at $t \approx 8$ hours. The variation shrinks to zero during this phase because the number of recordings $n$ reduces later in the night when the signal is lost.
%

As discussed in \sect{sect:radiotrack}, the movement is in two distinct phases, a dispersal followed by a return to the roost. In the next section various models for each phase will be compared. Although the data is discrete and stochastic in nature, we will seek to model the underlying probability distribution of the ensemble dynamics that is continuous in both space and time. We will use partial differential equations to analytically describe the evolution of the probability distribution.

\section{Phase 1: dispersal} \label{phase1}

Diffusion models are widely used to model animal movement, specifically dispersal, for a number of
species \cite{Ovaskainen2016}. Diffusion describes the movement of matter due to a random walk process from a region of high concentration to a region of low concentration. In this case, we wish to describe the dispersal  during phase 1 of movement as bats fly away from the roost (a region of high concentration) to the surrounding areas (regions of low concentration).  It is commonly accepted that bats tend to remain within an area around the roost known as the Core Sustenance Zone, and will forage within this area for the majority of the night \cite{CSZ}. As a result, a diffusion model on a bounded domain is considered here. First we will consider a one-dimensional diffusion model before extending to a two dimensional model in polar coordinates.
 %NOTE 1/2D models
 %As discussed in \chap{chap:intro}, Greater Horseshoe bats tend to move along linear features whilst foraging, and as such we will consider a one dimensional diffusion model in which bats are confined to moving along lines, as well as in a two dimensional diffusion model such that bats are able to travel anywhere within the domain.

 \subsection{A diffusion model in one dimension}

First we will consider diffusion on a one-dimensional domain, $\Omega \subset \mathbb{R}$, where $R$ is the size of the domain and the spatial coordinate is bounded, $x \in [0,R]$.  The probability density $\phi(x,t)$ of finding a bat at position $x$ at time $t$ is given by
 \begin{equation}
   \D{\phi(x,t)}{t} = D \DD{\phi(x,t)}{x} ,
   \label{eqn:diffusion_cartesian}
 \end{equation}
 where the diffusion coefficient, $D$, is a positive constant and quantifies the rate of spread. The boundary conditions,
\begin{align}
\D{\phi(x=0,t)}{x} &= 0, \\
\D{\phi(x=R,t)}{x} &= 0,
\label{eqn:BC1d}
\end{align}
specify zero-flux across the boundary, such that bats cannot enter or leave the domain. The initial condition,
\begin{equation}
\phi(x = 0) = \delta(0),
\label{eqn:IC1d}
\end{equation}
specifies that all bats begin the night at the roost before moving away to begin foraging. Theoretically, the boundary and initial conditions are not consistent. However, any computational solution requires a discretisation of the domain, and therefore the initial condition will be represented by an approximation of the delta function. Therefore, the computational solution will converge quickly to a stable solution. The roost is placed at $x=0$ to mimic the behaviour of diffusion in polar coordinates as we will be extending to polar coordinates later.
\subsection{Mean squared displacement in the one dimensional diffusion model} \label{msd1d}
The relationship between the expected mean squared displacement (MSD) and time $t$
can be calculated using moments of the probability density $\phi$. Although we can solve \eqn{eqn:diffusion_cartesian} direction, we will instead calculate the MSD through calculating the moments of the equation, as this is more useful when considering trajectory data. The $n$th moment is
\begin{equation}
\left<x^n\right> = \int_0^R x^n \phi(x,t) dx .
\label{eqn:moments_1d}
\end{equation}
The MSD is the 2nd moment, and is given by
\begin{equation}
\left<x^2\right> = \int_0^R x^2 \phi(x,t) dx .
\label{eqn:MSD_1d}
\end{equation}
Taking the time derivative of both sides and substituting $\D{\phi(x,t)}{t}$ from the diffusion \eqn{eqn:diffusion_cartesian},
\begin{align}
\frac{d}{dt} \left<x^2\right> &= \frac{d}{dt}\int_0^x x^2 \phi(x,t) dx ,\nonumber\\
                           &= \int_0^{R}  x^2 \D{\phi}{t} dx ,\nonumber\\
                           &= \int_0^{R} D \underbrace{\left[ x^2 \D{ \phi}{ x} \right]_0^{R}}_{=0} - D \int_0^{R} 2x \D{\phi}{x} dx, \nonumber\\
                            &= -2D \left[ x \phi \right]_0^{R} + 2D \int_0^{R} \phi dx , \nonumber\\
&= -2DR\phi(R,t) + 2D ,
\label{eqn:diffusion_1}
\end{align}
and therefore, integrating with respect to time gives
\begin{equation}
\left<x^2\right> = 2D \left( t - \int_0^t R \phi(R,\tau) d \tau \right).
\label{eqn:diffusion_msd1d}
\end{equation}
Over short timescales, $\phi(R,t) \approx 0$, since the probability of reaching the boundary over a short period of time is small due to the initial condition. Therefore, over a short timescale, whilst
\begin{equation}
\sqrt{2Dt} \ll R,
\end{equation}
the expected MSD for diffusion in one dimension is directly proportional to time,
\begin{equation}
\left<x^2\right> \approx 2Dt.
\label{eqn:diffusion_short1d}
\end{equation}
As noted in \sect{sect:radiotrack} the MSD in \fig{fig:MSD} is indeed linear during phase 1, and therefore consistent with a diffusion model.

An expression for the mean squared displacement over a long timescale can also be derived using \eqn{eqn:diffusion_msd1d}. Over long timescales, we expect the probability density to be uniformly spread across the domain,
\begin{equation}
\phi(x,t) = \frac{1}{R} .
\end{equation}
Substituting this into \eqn{eqn:MSD_expectation} gives
\begin{equation}
\left<x^2\right> = \frac{R^2}{3}
\label{eqn:diffusion_long1d}
\end{equation}
and therefore the mean squared displacement is constant over long timescales.

\begin{figure} [t]
    \centering
        \includegraphics[width=0.6\textwidth]{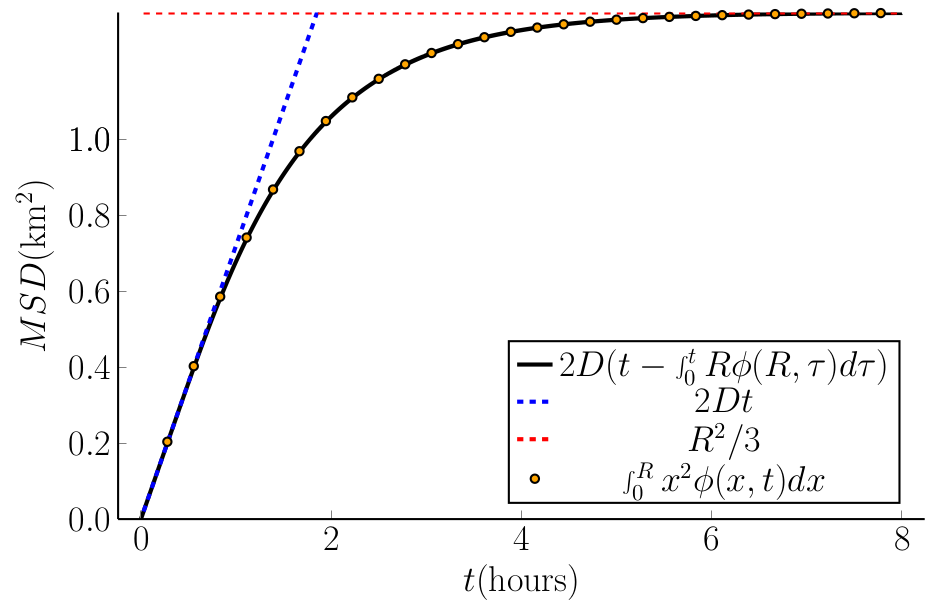}
        \caption{The expected MSD for a 1D diffusion model from \eqn{eqn:diffusion_msd1d} using a numerical solution to the diffusion equation with $D=100$m\textsuperscript{2}s\textsuperscript{-1}. The expectation value calculated using \eqn{eqn:MSD_1d} is shown along with the short and long timescale expressions in \eqn{eqn:diffusion_short1d} and \eqn{eqn:diffusion_long1d} for comparison.}
        \label{fig:expectedmsd1d}
\end{figure}

The expected MSD is plotted in \fig{fig:expectedmsd1d}, using a numerical solution to the
diffusion equation, \eqn{eqn:diffusion_cartesian}, solved using \texttt{DifferentialEquations.jl} \cite{DifferentialEquations}. The plot shows an initial linear section which tends to a constant value, consistent with \eqns{eqn:diffusion_short1d}{eqn:diffusion_long1d}.

 \subsection{A discretised diffusion model} \label{1ddiscrete}

\begin{figure} [t]
     \centering
         \includegraphics[width=0.5\textwidth]{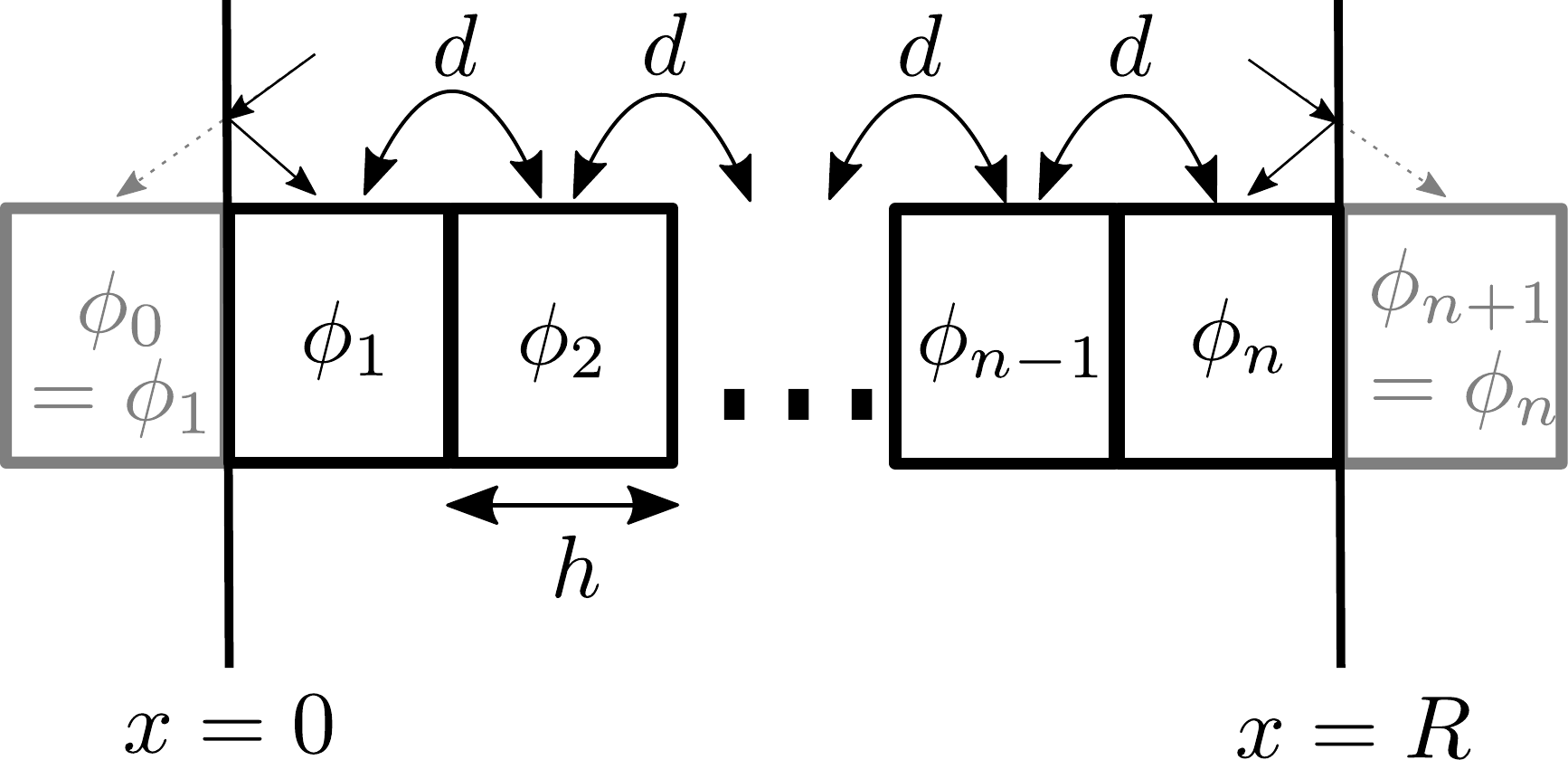}
         \caption{A diagram to illustrate the movement of probability density between boxes in the discretised diffusion model. Diffusion between boxes is represented by $d$ and the probability density in each box $i$ is denoted by $\phi_i$.}
     \label{fig:diffusion_diagram1d}
 \end{figure}

 The diffusion equation on a bounded domain can be solved using a discretised ODE description \cite{woolley2011stochastic}. The discretised formulation discussed here will also provide a framework for the models used for phase 2 of movement. The domain can be discretised into $N$ boxes, each of length
 $h=R/N$. The probability density in each box $i$ at $x=x_i$ is denoted by $\phi_i$, and evolves over time according to the diffusion process. A diagram of the motion is shown in \fig{fig:diffusion_diagram1d}. A finite difference approximation is used to describe the movement of probability density between boxes. The central difference approximation to the second order derivative at box $i$ is given by
 \begin{equation}
  \at{\DD{\phi}{x}}{x=x_i}= \frac{\phi_{i-1}-2\phi_i+\phi_{i+1}}{h^2}, \label{eqn:diffusion_discrete1di}
 \end{equation}
 where $\phi_i=\phi(x_i)$ and $\phi_{i \pm 1}=\phi{x_i \pm h}$.

 The equation for box 1 at $x = 0$ is
 \begin{equation}
 \frac{d\phi_1}{dt} = \frac{D}{h^2}(\phi_{0}-2\phi_1 +\phi_{2}).
 \end{equation}
 However, due to the reflective boundary condition in \eqn{eqn:BC1d}, any bats in a trajectory that would pass through the boundary are reflected back in the direction of the roost. Bats that would pass from box 1 to an imaginary box 0 are instead reflected back, and therefore the value of $\phi$ in the imaginary box 0 is the same as in box 1, $\phi_0$ = $\phi_1$, and
 \begin{equation}
 \frac{d\phi_1}{dt} = \frac{D}{h^2}(\phi_{2}- \phi_1).
         \label{eqn:box_1}
 \end{equation}
 Similarly, from \eqn{eqn:diffusion_discrete1di}, for box $n$ at $x=R$,
 \begin{equation}
 \frac{d\phi_n}{dt} = \frac{D}{h^2}(\phi_{n-1}-2\phi_n +\phi_{n+1}).
 \end{equation}
 Due to the reflective boundary condition between box $n$ and box $n+1$, $\phi_{n+1} = \phi_n$, and
 \begin{equation}
 \frac{d\phi_n}{dt} = \frac{D}{h^2}(\phi_{n-1}-\phi_n).
         \label{eqn:annulus_n1d}
 \end{equation}

\begin{figure} [t]
     \centering
         \includegraphics[width=0.8\textwidth]{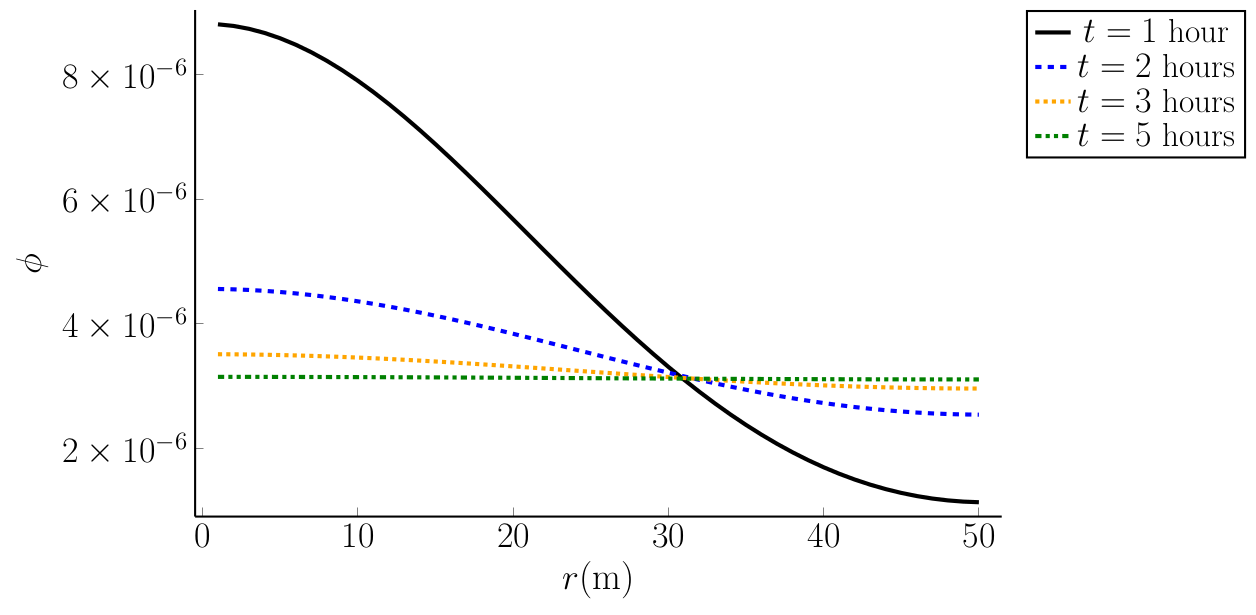}
         \caption{The value of $\phi(r)$ at position $r$ for a 1D discretised diffusion simulation with parameters $N = 100$, $R = 2000\mathrm{m}$ and $D = 100\mathrm{m^2s^{-1}}$ and initial condition given by \eqn{eqn:discrete_diffusion_IC} after $t = $ 1 hour, 2 hours and 5 hours.}
     \label{fig:discretised_phi}
 \end{figure}

 Collecting \eqnto{eqn:diffusion_discrete1di}{eqn:annulus_n1d} and substituting $d = D/h^2$, the set of equations describing the full system is
 \begin{equation}
 \frac{d\phi_i}{dt} = \begin{cases}
 		d(\phi_i - \phi_{i+1}), & \text{for } i = 1, \\
 		d(\phi_{i-1}-2\phi_i +\phi_{i+1}), & \text{for } 2 \leq i \leq N-1, \\
 		d(\phi_{i-1}-\phi_i), & \text{for } i = N ,
 		\end{cases}
         \label{eqn:discrete_diffusion}
 \end{equation}
 where the discretised diffusion coefficient is given by
 $d = D/h^2$.
  The initial condition corresponding to \eqn{eqn:IC1d} means that probability density is concentrated in the first box,
 \begin{equation}
 \phi_i(0) = \begin{cases}
 		\frac{1}{h}, & \text{for } i = 1, \\
 		0, & \text{for } 2 \leq i \leq N. \\
 		\end{cases}
         \label{eqn:discrete_diffusion_IC}
 \end{equation}
 A diagram illustrating the spread of probability density due to the diffusion process is shown in \fig{fig:diffusion_diagram1d}. This generates a system of $N$ ODEs describing motion over
 the domain at each time step and which can be solved using a numerical ODE solver. The equations for $i=1$ and $i=N$ correspond to reflective, zero-flux boundary conditions.  The result of a simulation with $N = 100$ boxes in a domain of
 length $R = 1000$m and diffusion coefficient $D = 100\mathrm{m^2s^{-1}}$ is shown in
 \fig{fig:discretised_phi}. The diffusion process spreads the probability density out from the left side of the domain, where $\phi$ is high and tends to homogenise the probability density across the domain over time. After 5 hours, the probability density $\phi$ is evenly spread throughout the domain and the probability distribution is eventually uniform.

\subsection{Diffusion in two dimensions}

Next we will consider a diffusion model in two dimensions to
describe dispersal during phase 1 of movement. The third dimension is not
included as height is not measured in the radio tracking survey, and it is not
needed to describe landscape use.
If the roost is at $(x_0,y_0)$ and bats leave the roost at time $t =0$,
the 2D diffusion equation describes the probability density $\phi(x,y,t)$ of
finding a bat at position $(x,y)$ at time $t$,
\begin{equation}
  \D{\phi(x,y,t)}{t} = D \nabla^2 \phi(x,y,t) ,
  \label{eqn:diffusion_cartesian2d}
\end{equation}
where $\nabla^2$ is the Laplacian, $D$ is the diffusion coefficient, a positive constant that quantifies the
 rate of spread. The Core Sustenance Zone is denoted by $\Omega \subset \mathbb{R}^2$ and modelled as a disk of radius $R$ centred around the roost. Therefore we will consider the diffusion equation in polar coordinates,
 \begin{equation}
 \D{ \phi(r,t)}{t} = \frac{D}{r} \D{}{ r} \left( r \D{\phi(r,t)}{r} \right),
 \label{eqn:diffusion_polar2d}
 \end{equation}
 where $r$ is the distance from the roost, given by $r=\sqrt{(x-x_0)^2 +
 (y-y_0)^2}$. Since the domain is symmetric, $\phi$ is only dependent on $r$ and not
  on the angle. The initial condition,
 \begin{equation}
 \phi(r = 0) = \delta(0),
 \label{eqn:IC2d}
 \end{equation}
specifies that all bats begin the night at the roost at position $r=0$ before moving away at time $t$ to begin foraging. The boundary condition,
\begin{equation}
\D{\phi(r=R,t)}{r} = 0,
\label{eqn:BC}
\end{equation}
specifies zero-flux across the boundary such that bats cannot enter or leave the boundary.

\subsection{Mean squared displacement in the two dimensional diffusion model} \label{msd2d}
The relationship between the expected mean squared displacement (MSD) and time $t$
can be calculated using moments of the probability density $\phi$, which are defined similarly to \eqn{eqn:moments_1d},
\begin{equation}
\left<r^2\right> = \int_{\Omega}r^2 \phi(r,t) d\omega ,
\label{eqn:MSD_expectation}
\end{equation}
where $\omega = (r,\theta) \subset \Omega$. Taking the time derivative of both
sides and substituting $\D{\phi(r,t)}{t}$  from \eqn{eqn:diffusion_polar2d},
\begin{align}
\frac{d}{dt} \left<r^2\right> &= \frac{d}{dt}\int_{\Omega}r^2 \phi(r,t) d\omega , \nonumber\\
                           &= \int_0^{2\pi} \int_0^{R} r^3 \D{\phi}{t} dr d\theta ,\nonumber\\
                            &= \int_0^{2\pi} \int_0^{R} r^3 \frac{D}{r} \D{}{r } \left( r \D{ \phi}{ r}\right) dr d\theta , \nonumber\\
                            &= \int_0^{2\pi} \int_0^{R} -2r^2D \D{ \phi}{r}dr d\theta , \nonumber\\
                            &= - 4\pi R^2D \phi(R,t) + 4D \int_{\Omega} \phi d\omega ,
\end{align}
and therefore,
\begin{equation}
\frac{d}{dt} \left<r^2\right>  = 4D( 1- \pi R^2 \phi(R,t)) .
\end{equation}
Integrating with respect to time gives
\begin{equation}
\left<r^2\right> = 4D \left( t - \pi R^2 \int_0^t \phi(R,\tau) d \tau \right).
\label{eqn:diffusion_msd2d}
\end{equation}
Over short timescales, $\phi(R,t) \approx 0$, since the probability of a reaching the boundary over a short period of time is small due to the initial condition. Therefore, over a short timescale, while
\begin{equation}
t \ll \pi R^2,
\end{equation}
the expected MSD for diffusion is directly proportional to time,
\begin{equation}
\left<r^2\right> \approx 4Dt.
\label{eqn:diffusion_short2d}
\end{equation}
An expression for mean-squared displacement over a long timescale can also be derived using \eqn{eqn:MSD_expectation}. Over long timescales, we expect the probability density to be uniformly spread across the domain,
\begin{equation}
\phi(r,t) = \frac{1}{\pi R^2} .
\end{equation}
Substituting this into \eqn{eqn:MSD_expectation} gives
\begin{align}
\left<r^2\right> &= \frac{1}{\pi R^2}\int_{0}^{2\pi} \int_{0}^{R} r^3 dr d\theta \nonumber \\
&= \frac{1}{2} R^2,
\label{eqn:diffusion_long2d}
\end{align}
and therefore the mean squared displacement is constant over long timescales.

\begin{figure} [t]
    \centering
        \includegraphics[width=0.6\textwidth]{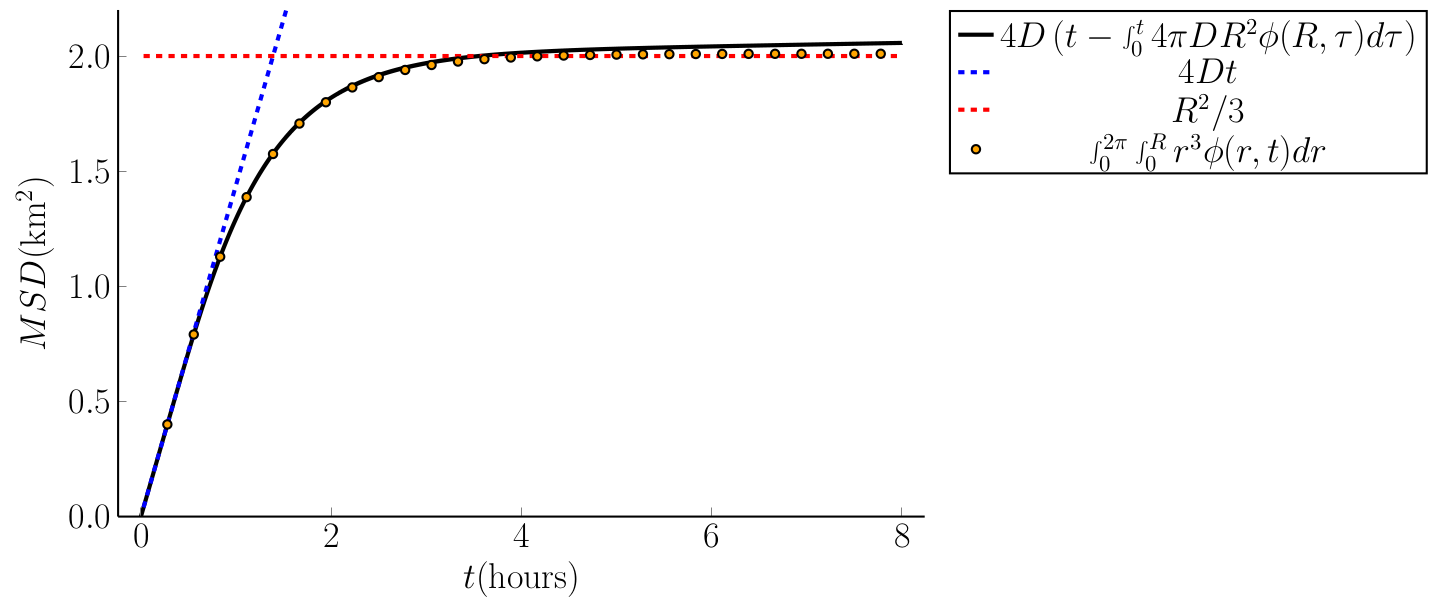}
        \caption{The expected MSD for the polar diffusion model defined by \eqn{eqn:diffusion_msd2d} using a numerical solution to the diffusion equation with $D=100$m\textsuperscript{2}s\textsuperscript{-1}. The expectation value calculated using \eqn{eqn:MSD_expectation} is shown along with the short and long timescale expressions in \eqn{eqn:diffusion_short2d} and \eqn{eqn:diffusion_long2d} for comparison.}
        \label{fig:expectedmsd2d}
\end{figure}

The expected MSD is plotted in \fig{fig:expectedmsd2d}, using a numerical solution to the
diffusion equation, \eqn{eqn:diffusion_polar2d}, solved using \texttt{DifferentialEquations.jl} \cite{DifferentialEquations}. The plot shows an initial linear section which tends to a constant value, consistent with \eqns{eqn:diffusion_short2d}{eqn:diffusion_long2d}.

\subsection{A discretised ODE solution to the two dimensional diffusion model}

\begin{figure} [t]
\centering
    \includegraphics[width=0.3\textwidth]{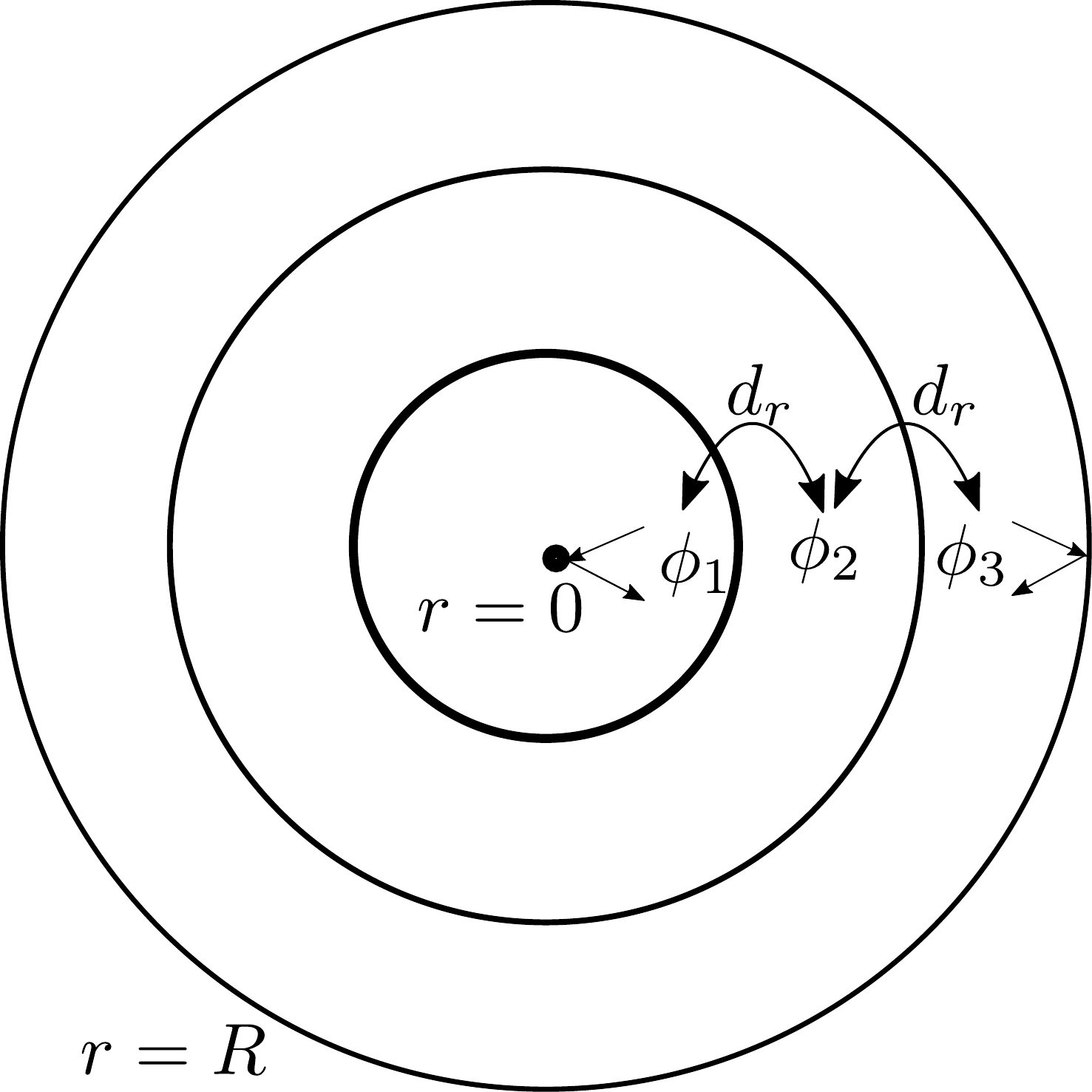}
    \caption{A diagram of the discretised diffusion process. Each box represents an annular region around the roost centre. Here we illustrate the domain discretised into 3 sections, but the diagram extends analogously to any number of compartments. The probability density shifts between boxes due to the diffusion process, denoted by $d_r$, where $d_r = D \left(\DD{\phi(r,t)}{r} + \frac{1}{r}\D{\phi(r,t)}{r}\right)$. Reflective boundary conditions ensure that probability density never leaves the domain.}
\label{fig:2d_diffusion_diag}
\end{figure}

The polar diffusion equation can be discretised using central difference approximations in the same way as for one dimensional Cartesian PDEs \cite{mori2015numerical,galeriu2004modeling,britt2010compact}, and this discretisation will be used to solve the equation in a circular domain. For a circular domain in polar coordinates $\Omega = [0,R] \times [0, 2\pi]$, the domain can be discretised into $N$ annuli, each of
width $h=R/N$. Due to the assumption that the domain is symmetric, we will not consider angular movement around the annuli, and we can reduce the problem to an advection-diffusion problem in only one dimension. The probability density in each annulus $i$ is denoted by $\phi_i$, and evolves over time according to the diffusion process.  The distance from the origin to the inner edge of annulus $i$ is given by $r_i$. A diagram of the motion is shown in \fig{fig:2d_diffusion_diag}. Central difference approximations to \eqn{eqn:diffusion_polar2d} are used to describe the movement of probability density between annuli. First we will expand the differential in \eqn{eqn:diffusion_polar2d},
\begin{equation}
\D{ \phi(r,t)}{t} = D \left(\DD{\phi(r,t)}{r} + \frac{1}{r}\D{\phi(r,t)}{r} \right).
\label{eqn:diffusion_polar2d_expand}
\end{equation}
The central difference approximation to the first and second order derivatives at annulus $i$ are given by
\begin{align}
 \at{\D{\phi}{r}}{r=r_i}&= \frac{1}{r_i}\frac{\phi(r_i+h)-\phi(r_i-h)}{2h}, \\
 \at{\DD{\phi}{r}}{r=r_i}&= \frac{\phi(r_i-h)-2\phi(r_i)+\phi(r_i+h)}{h^2}. \label{eqn:diffusion_discretei}
\end{align}
The central difference approximation to \eqn{eqn:diffusion_polar2d_expand} is then
\begin{equation}
 \at{\D{\phi}{t}}{r=r_i} = D \left(\frac{\phi(r_i-h)-2\phi(r_i)+\phi(r_i+h)}{h^2} + \frac{1}{r_i} \frac{\phi(r_i+h)-\phi(r_i-h)}{2 h} \right).
 \label{eqn:diffusion_polar_cd}
\end{equation}
The distance $r_i$ at annulus $i$ is given by
\begin{equation}
r_i = i h ,
\end{equation}
and \eqn{eqn:diffusion_polar_cd} can therefore be written as
\begin{equation}
\D{\phi(r_i)}{t} = \frac{D}{h^2} \left(\phi(r_i-h)-2\phi(r_i)+\phi(r_i+h)\right) + \frac{D}{2ih^2}\left(\phi(r_i+h)-\phi(r_i-h) \right).
\label{eqn:diffusion_polar_final}
\end{equation}
Changing notation for the discretised version gives
\begin{equation}
\frac{d\phi_i}{dt} = \frac{D}{h^2}(\phi_{i-1}-2\phi_i +\phi_{i+1}) + \frac{D}{2ih^2} (\phi_{i+1}-\phi_{i}).
        \label{eqn:discrete_diffusion_i}
\end{equation}
The equation for annulus 1 at $r=0$ is
\begin{equation}
\frac{d\phi_1}{dt} = \frac{D}{h^2}(\phi_{0}-2\phi_1 +\phi_{2}) + \frac{D}{2h^2} (\phi_{2}-\phi_{1}).
\end{equation}
However, due to the reflective boundary condition in \eqn{eqn:BC}, any bats in a trajectory that would pass through the boundary are reflected back in the direction of the roost. Bats that would pass from annulus 1 to an imaginary annulus 0 are instead reflected back, and therefore the value of $\phi$ in the imaginary annulus 0 is the same as in annulus 1, and $\phi_0$ = $\phi_1$.
\begin{equation}
\frac{d\phi_1}{dt} = \frac{D}{h^2}(\phi_{2}- \phi_1) + \frac{D}{2h^2} (\phi_{2}-\phi_{1}) .
        \label{eqn:annulus_1}
\end{equation}

Similarly, from \eqn{eqn:diffusion_discrete1di}, for annulus $n$ at $r=R$,
\begin{equation}
\frac{d\phi_n}{dt} = \frac{D}{h^2}(\phi_{n-1}-2\phi_n +\phi_{n+1}) + \frac{D}{2h^2} (\phi_{n+1}-\phi_{n}) .
\end{equation}
Due to the reflective boundary condition between annulus $n$ and annulus $n+1$, $\phi_{n+1} = \phi_n$, and
\begin{equation}
\frac{d\phi_n}{dt} = \frac{D}{h^2}(\phi_{n-1}-\phi_n) + \frac{D}{2h^2} (\phi_{n+1}-\phi_{n-1}) .
        \label{eqn:annulus_n}
\end{equation}

Collecting \eqnto{eqn:discrete_diffusion_i}{eqn:annulus_n}, the set of equations describing the full system is
  \begin{equation}
  \frac{d\phi_i}{dt} = \begin{cases}
  		\frac{D}{h^2}(\phi_{i+1} - \phi_{i}) + \frac{D}{2h^2} (\phi_{i+1}-\phi_i), & \text{for } i = 1, \\
  		\frac{D}{h^2}(\phi_{i-1}-2\phi_i +\phi_{i+1}) + \frac{D}{2ih^2} (\phi_{i+1}-\phi_{i}), & \text{for } 2 \leq i \leq N-1, \\
  		\frac{D}{h^2}(\phi_{i-1}-\phi_i) + \frac{D}{2ih^2} (\phi_{i}-\phi_{i-1}), & \text{for } i = N .
  		\end{cases}
          \label{eqn:discrete_diffusion2d}
  \end{equation}
 The initial condition corresponding to \eqn{eqn:IC1d} means that probability density is concentrated in the first annulus at $t=0$,
\begin{equation}
\phi_i(0) = \begin{cases}
       \frac{1}{2\pi h}, & \text{for } i = 1, \\
       0, & \text{for } 2 \leq i \leq N. \\
       \end{cases}
        \label{eqn:discrete_diffusion_IC2d}
\end{equation}
%
%\begin{figure}
%\centering%
%    \includegraphics[width=0.8\textwidth]{figs/2d_diffusion.png}
%    \caption{The mean squared displacement for a discretised diffusion simulation with $N=1000$ particles in a bounded 2D domain for $D=65$ms\textsuperscript{-2} compared to the analytical result of $4Dt$ from \sect{msd2d}.}
%\label{fig:2d_diffusion}
%\end{figure}
%
\subsection{Comparison of one and two dimensional diffusion models}
\begin{figure}
\centering
    \includegraphics[width=0.6\textwidth]{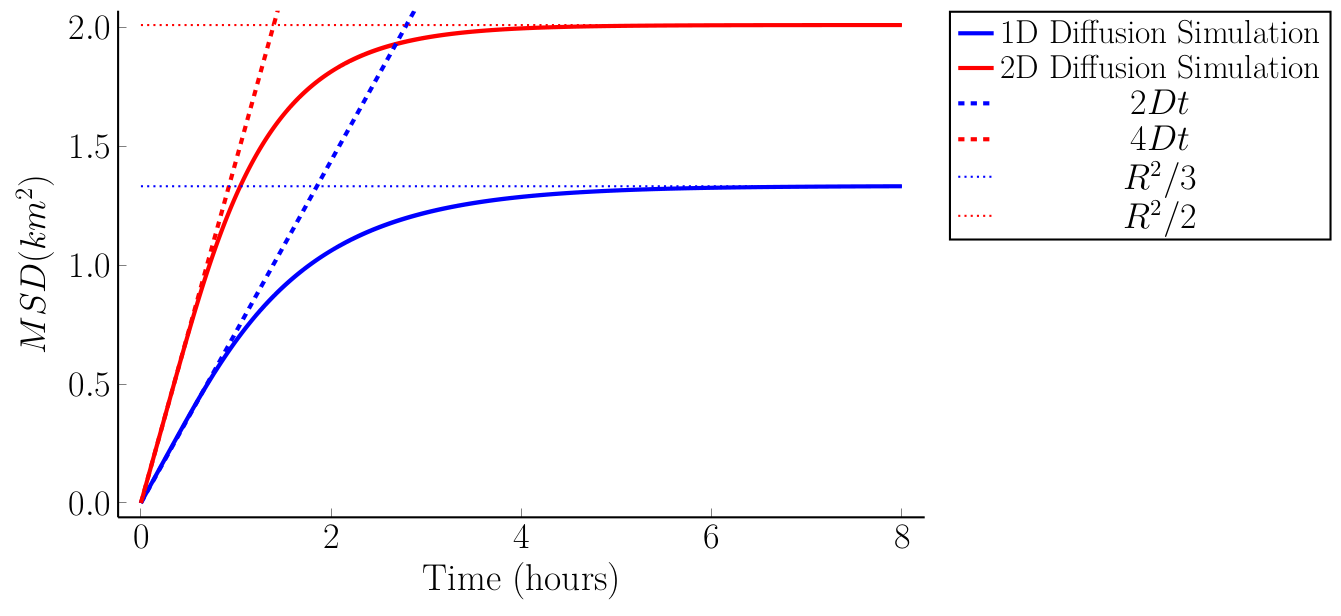}
    \caption{A comparison of the mean squared displacement for discretised diffusion simulations in bounded one and two dimensional domains, both with $N=1000$ particles and $D=100$ms\textsuperscript{-2}, compared to the analytical results from \sect{msd1d} and \sect{msd2d}.}
\label{fig:1d2d_diffusion}
\end{figure}

Diffusion simulations in 1D and 2D were simulated, each with $N=1000$ particles and diffusion coefficient $D=100$m\textsuperscript{2}s\textsuperscript{-1}, using \texttt{DifferentialEquations.jl} \cite{DifferentialEquations} to validate the model. The mean squared displacement for each is shown in \fig{fig:1d2d_diffusion}. The mean squared displacements for both discretised simulations are initially linear and consistent with the analytical results calculated in \sect{msd1d} and \sect{msd2d}. The curve then flattens and tends towards a constant value as expected.

\section{Phase 2: return to roost}

The diffusion model described in \sect{phase1} explains the initial dispersal in phase 1, however it cannot explain the decrease in MSD for phase 2. For the second phase of movement, two different models are considered, a convection-diffusion diffusion model and a diffusion model on a shrinking domain. Convection-diffusion models describe movement under the influence of two processes, diffusion and convection. The diffusion component corresponds to dispersal as discussed in \sect{phase1}, whilst the convection component describes a drift in a particular direction, or towards a specific location. These models are commonly used to model animal movement in response to external factors, for example a drift towards patches of high resources or away from predators. However we will show here that a diffusion model in a shrinking domain provides a more accurate description of bat movement whilst foraging.

\subsection{A convection-diffusion model in two dimensions} \label{sect:convection}

During phase 2 of movement, the MSD is decreasing as bats return towards a point, the roost location $(x_0,y_0)$ at $r=0$. We will first consider a  convection-diffusion model to describe this drift. The 2D symmetric convection-diffusion equation, in polar coordinates, is
\begin{equation}
  \D{\phi(r,t)}{t} = \frac{D}{r} \D{}{r}\left(r \D{\phi(r,t)}{r} \right) - \chi \D{\phi(r,t)}{r},
  \label{eqn:convection}
\end{equation}
where $D$ is the diffusion coefficient and $\chi$ is the convection coefficient. As bats are heading towards $r=0$, the convection component of \eqn{eqn:convection} describes a drift towards $r=0$. As bats undergo diffusive movement whilst dispersing from the roost for a time $T$ before their behaviour changes, $\chi$ is time dependent,
\begin{equation}
\chi (t) =  \begin{cases}
    0, & \text{for } t < T, \\
    \chi_0, & \text{for } t \ge T,
  \end{cases}
  \label{eqn:time_dependentchi}
\end{equation}
where $\chi_0$ is a positive constant. When $t < T$, the convection term in \eqn{eqn:convection} is zero, and the equation reduces to a polar diffusion equation, as in \eqn{eqn:diffusion_polar2d}. For $t > T$, \eqn{eqn:convection} becomes
\begin{equation}
  \D{\phi(r,t)}{t} = \frac{D}{r} \D{}{r}\left(r \D{\phi(r,t)}{r} \right) - \chi_0 \D{\phi(r,t)}{r}.
\end{equation}

\subsection{A discretised convection-diffusion model} \label{sect:convectiondiscrete}

 \begin{figure} [ht]
     \centering
         \includegraphics[width=0.3\textwidth]{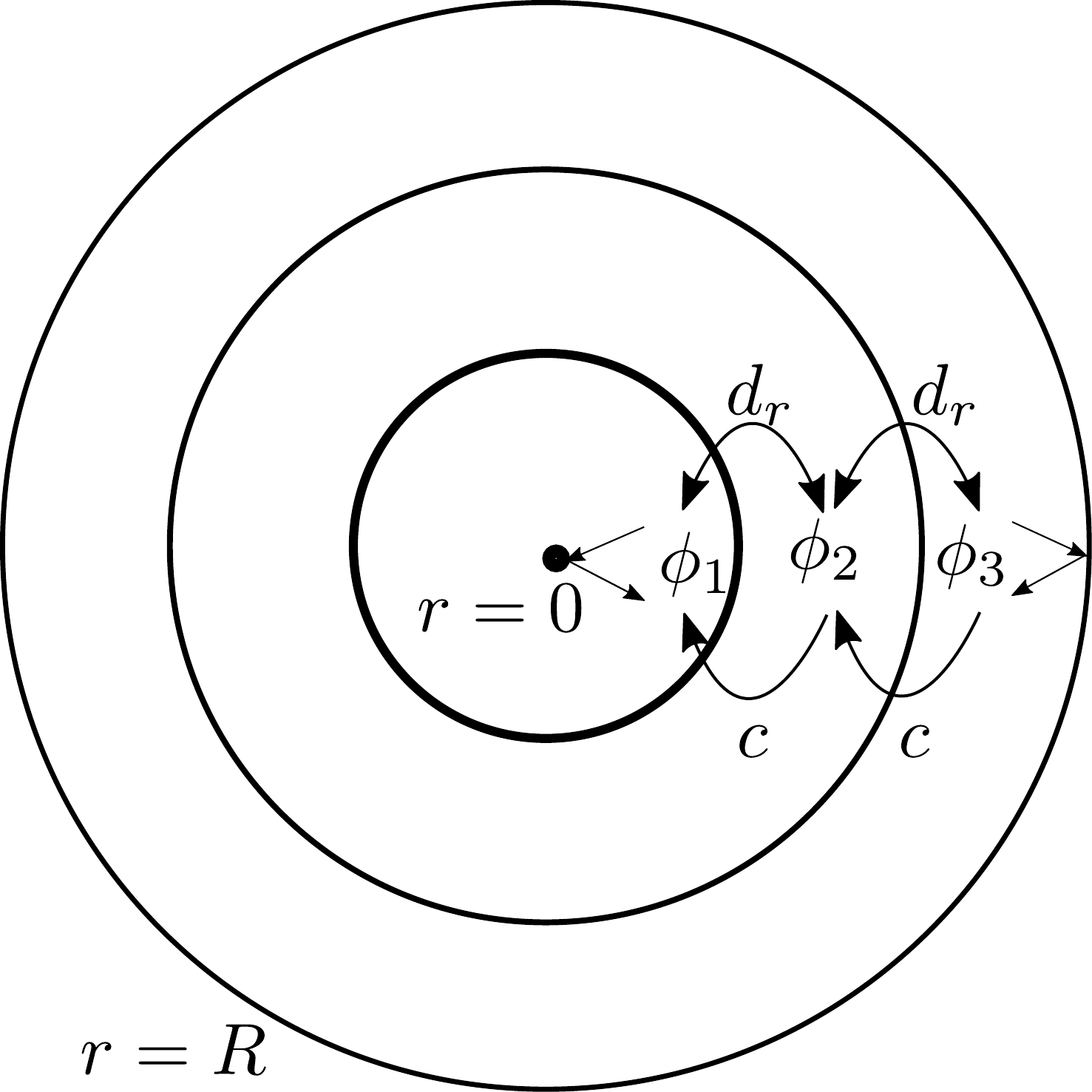}
         \caption{A diagram to illustrate the movement of probability density between annuli in the discretised convection-diffusion model. Here we illustrate the domain discretised into 3 sections, but the diagram extends analogously to any number of compartments. As there is no angular component to this movement, the problem is reduced to one radial dimension. Diffusion between annuli is represented by $d$ and the drift due to convection is represented by $c$. The probability density in each annulus $i$ is denoted by $\phi_i$. Here we illustrate the domain discretised into 5 sections, but the diagram can extend analogously to any number of components. }
     \label{fig:convection_diffusion_diag}
 \end{figure}

The convection-diffusion model will be solved using a discretised ODE model, as with the diffusion model in \sect{phase1}. The domain $\Omega$ of length $R$ is discretised into $N$ annuli, each of length $h = R/N$. The probability density in each annulus $i$ is denoted by $\phi_i$ and evolves over time according to the diffusion and convection processes, as illustrated in \fig{fig:convection_diffusion_diag}. The diffusion component is unchanged from \sect{phase1}, however this time there is an additional convection component pushing bats away from the domain boundary and back towards the roost at $r=0$. The convection process shifts probability density towards the left, towards annulus $i = 1$, and the discretised equations are
\begin{equation}
\frac{d\phi_i}{dt} = \begin{cases}
		d(\phi_{i+1} - \phi_i) + \frac{d}{2} (\phi_{i+1}-\phi_i) - c \phi_{i+1}, & \text{for } i = 1, \\
		d(\phi_{i-1}-2\phi_i +\phi_{i+1}) + \frac{d}{2i} (\phi_{i+1}-\phi_{i-1}) - \frac{c}{i}(\phi_{i+1}-\phi_{i}), & \text{for } 2 \leq i \leq N-1, \\
		d(\phi_{i-1}-\phi_i)  + \frac{d}{2i} (\phi_{i}-\phi_{i-1}) - \frac{c}{i}\phi_{i}, & \text{for } i = N.
		\end{cases}
        \label{eqn:discrete_convection}
\end{equation}
The discretised diffusion coefficient is $d = D/h^2$ as before and $c=\chi/h$ is the discretised convection coefficient. The initial condition is the state of the system after diffusion for time $T$.

\begin{figure}
     \centering
     \begin{subfigure}[b]{0.48\textwidth}
         \centering
         \includegraphics[width=\textwidth]{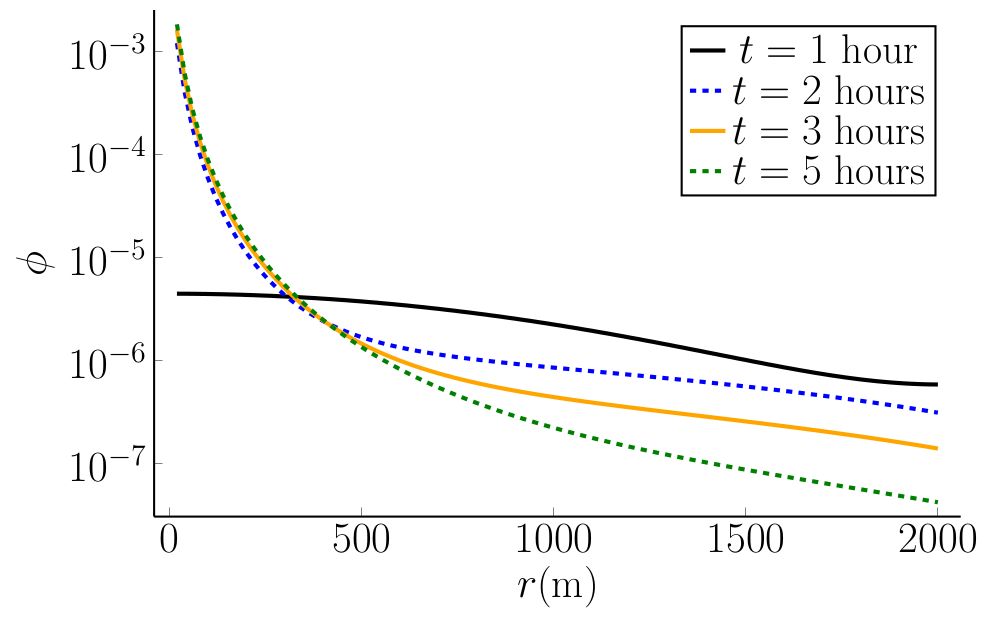}
         \caption{The probability density $\phi$ after $t = 1$, 1.5, 2 and 5 hours.}
         \label{fig:convection2dphi}
     \end{subfigure}
     \hfill
     \begin{subfigure}[b]{0.48\textwidth}
         \centering
         \includegraphics[width=\textwidth]{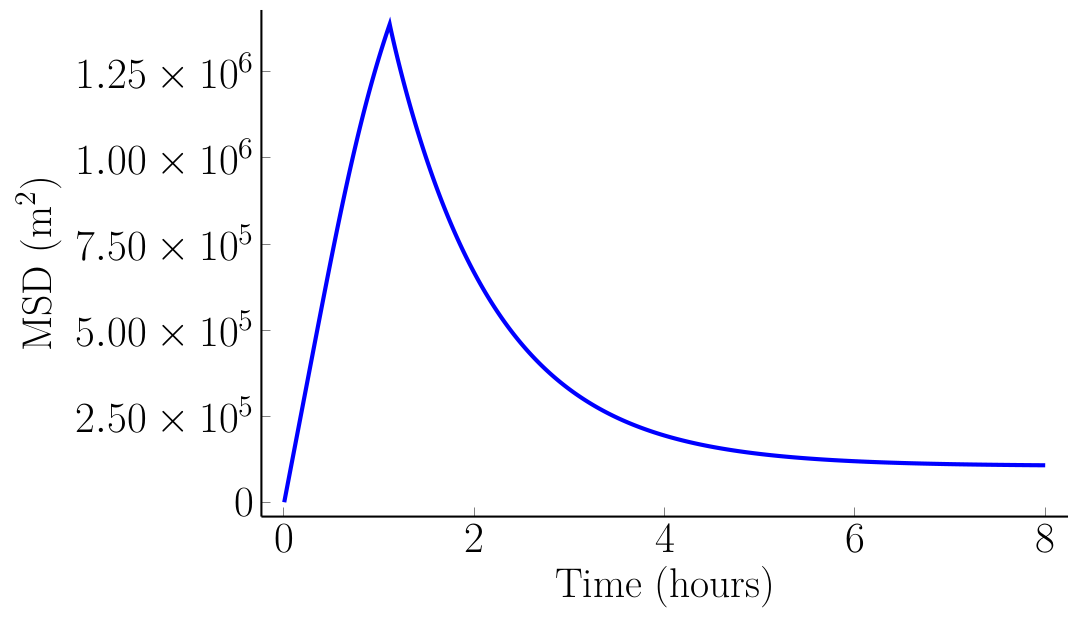}
         \caption{The mean squared displacement for $0 \leq t \leq 8$ hours. }
         \label{fig:convection2dmsd}
     \end{subfigure}
     \caption{The results of a convection-diffusion simulation with parameters $R = 2000$m, $N = 100$,
     $D = 100\mathrm{m^2s^{-1}}$, $\chi_0 = - 15\mathrm{ms^{-1}}$ and $T = 4000$ seconds.}
     \label{fig:convection2d}
     \end{figure}

The model was simulated using \texttt{DifferentialEquations.jl} \cite{DifferentialEquations}, using a time dependent convection coefficient as in \eqn{eqn:time_dependentchi} with $T = 4000$ seconds. The
simulation was run with a domain of length $R = 2000$m, split into $N = 100$
annuli. The diffusion coefficient was $D = 65\mathrm{m^2s^{-1}}$, and the convection coefficient
was $\chi =  - 15\mathrm{ms^{-1}}$. The results
of this simulation are shown in \fig{fig:convection2d}. The probability density $\phi$ is shown in \fig{fig:convection2dphi}. The convection process pushes $\phi$ uniformly in the direction of the drift, towards the left side of the domain, whereas the diffusion tends to spread $\phi$ across the domain. As bats reach the roost at $r=0$, they enter the roost and stop moving. This is analogous to an absorbing boundary at $r=0$. The boundary acts as a barrier stopping $\phi$ from moving any further and $\phi$ collects at the boundary. The movement towards the edge slows as $t$ increases because as each bat enters the roost, it stops moving, and eventually the system reaches a steady state when diffusion and convection are balanced.

The MSD for the same convection-diffusion simulation is shown in \fig{fig:convection2dmsd}. The convection-diffusion simulation shows the initial rapid dispersal expected from the diffusion model, however the shape of the curve for phase 2 is clearly inconsistent with the radio tracking data. The convection-diffusion simulation yields a convex curve, and the decrease in MSD slows with time as bats return to the roost and stop contributing to the movement. The MSD never reaches 0 because the diffusion term acts to spread bats out whilst the convection term is pushing them back towards the roost, and these eventually balance, without the colony returning to the roost.

\begin{figure} [h]
    \centering
        \includegraphics[width=0.7\textwidth]{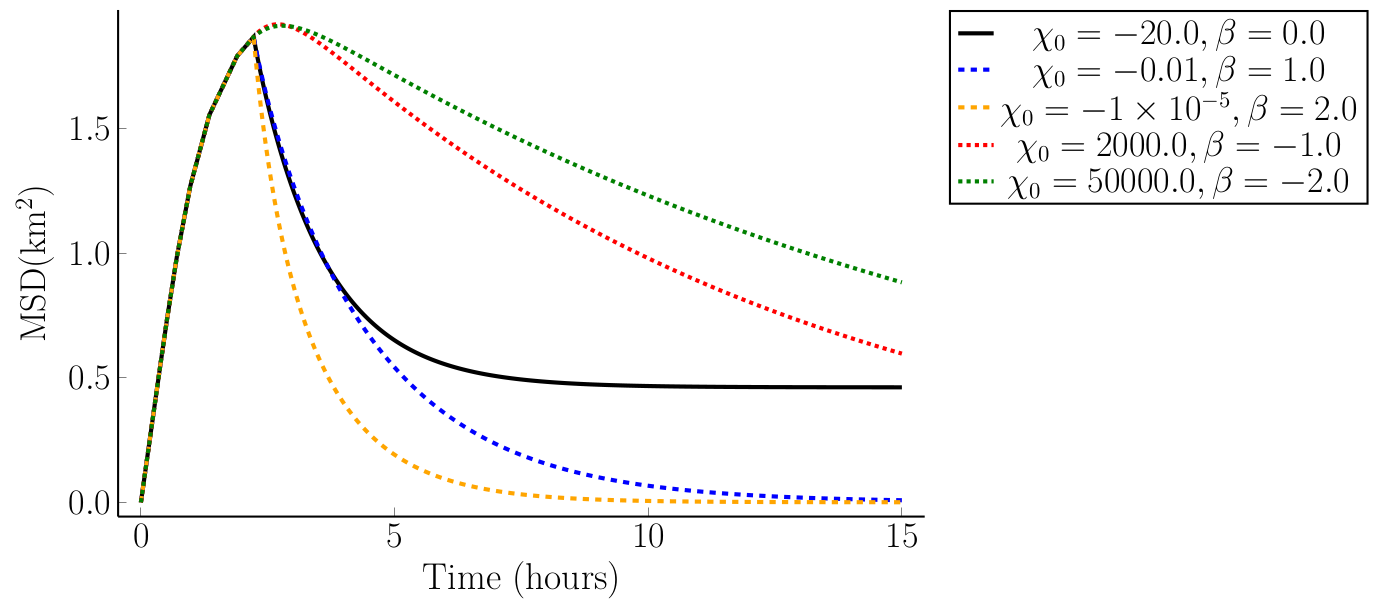}
        \caption{The mean squared displacement for spatially dependent convection-diffusion models with convection coefficient $\chi$ of the form given in \eqn{eqn:r_dependentchi}. The simulation parameters are $R = 2000$m, $N = 100$,
        $D = 100\mathrm{m^2s^{-1}}$ and $T = 4000$ seconds.}
    \label{fig:rdependentphi}
\end{figure}

In \fig{fig:MSD}, the curve is concave because the MSD decreases slowly at first, when bats are most spread out and furthest from the roost, and the rate of decrease increases with time. One possible model that may provide a concave curve in the MSD could be to
use a convection-diffusion model in which the convection coefficient is spatially dependent as well as time dependent. In this case, we can engineer the spatial dependence such that when bats are far from the roost, they drift slowly, and the drift speeds up as they get closer to the roost. In this case, bats drift back to the roost at a rate dependent on their distance from the roost,
\begin{equation}
\chi (r,t) =  \begin{cases}
    0, & \text{for } t < T, \\
    r^{\beta}\chi_0, & \text{for } t \ge T,
  \end{cases}
  \label{eqn:r_dependentchi}
\end{equation}
where $\beta$ is a constant. The results of simulations with $ -2 \leq \beta \leq 2$ are shown in \fig{fig:rdependentphi}. The plots show that for each value of the exponent $\beta$, the curves are convex rather than concave as diffusion eventually balances convection and bats stop moving once they reach the roost. When $\beta$ is negative, increasing $\chi$ pushes bats closer to the roost, however the value must be artificially inflated in order to overcome diffusion, and does not solve the convexity problem.

From \fig{fig:rdependentphi} we can see that a convection-diffusion model of this form cannot produce a concave curve because under this model bats stop moving when they reach the roost. In order to produce a concave curve consistent with the radio tracking data, the model cannot push all bats towards the roost at the same time, as those closest to the roost will always reach the roost first and stop moving. Next we will consider a model that solves this problem by selecting only the bats furthest from the roost to drift back towards the roost. We will first derive a stochastic model describing bat movement and then use it to inform a deterministic model.

 	\subsection{A stochastic diffusion model}

 	For simplicity, we will first derive a stochastic diffusion model in one
 	dimension, before extending to two dimensions. The solution to the unbounded one dimensional diffusion \eqn{eqn:diffusion_cartesian}, with the initial condition
    \begin{equation}
    \phi(x = 0) = \delta(0),
    \label{eqn:stochastic_IC}
    \end{equation}
   is given by
 	\begin{equation}
 \phi(x,t) = \frac{1}{\sqrt{4\pi Dt}}\exp \left(\frac{-x^2}{4Dt} \right) ,
 \label{eqn:diffusion_solution1d}
 	\end{equation}
 equivalent to a normal distribution with mean $\mu = 0$ and variance $\sigma^2 =
 2Dt$.

 To convert to a stochastic diffusion model, we can convert the deterministic
 diffusion equation to a stochastic differential equation with discrete time and
 continuous space \cite{ghoniem1985grid,roberts2002langevin}. We assume that bats
 are able to move freely through space, with the location recorded at regular
 time steps, with interval $\tau$. Given the initial condition in \eqn{eqn:stochastic_IC}, the probability density after one time step $\tau$ must be
 \begin{equation}
 \phi(x,t) = \frac{1}{\sqrt{4\pi D\tau}}\exp \left(\frac{-x^2}{4D\tau} \right) ,
 \end{equation}
 equivalent to a normal distribution with mean $\mu = 0$ and variance $\sigma^2 =
 2D\tau$. Therefore, for the first time step, each particle takes a step $dx$
 drawn at random from this distribution, $dx \sim \mathcal{N}(0,2D\tau)$. If each
 particle then takes a second step, drawn from the same distribution, the
 probability density at $x$ and $t = 2\tau$ is given by the sum of the
 probabilities that the particle reaches $x$ in 2 steps, the integral
 \begin{equation}
 \phi(x,t=2\tau) = \int_{-\infty}^{\infty} \frac{1}{4\pi D \tau} \exp \left(\frac{-\eta^2}{4D\tau} \right) \exp \left(\frac{-(x-\eta)^2}{4D\tau}\right) d\eta.
 \end{equation}
 The result of this integral is given by
 \begin{equation}
 \phi(x,t=2\tau) = \frac{1}{\sqrt{8\pi D\tau}} \exp \left(\frac{-x^2}{8\pi D\tau} \right).
 \end{equation}
 By induction, after $k$ steps, at time $k\tau$, each chosen from the same distribution, $dx \sim \mathcal{N}(0,2D\tau)$, the probability density at $x$ is
 \begin{equation}
 \phi(x,t=k\tau) = \frac{1}{\sqrt{4k\pi D\tau}} \exp \left(\frac{-x^2}{4k\pi D\tau} \right),
 \end{equation}
 equivalent to the distribution given by the solution to the diffusion equation at time $t=k\tau$, \eqn{eqn:diffusion_solution1d}.
 We can write this process as a stochastic differential equation,
 \begin{equation}
 dx_i = \rho_i,
 \label{eqn:SDE}
 \end{equation}
 where $\{\rho_i\}$ is a set of random numbers chosen from the normal distribution with zero mean and standard deviation $\sqrt{2D\tau}$, such that $\rho_i \sim \mathcal{N}(0,2D\tau)$. The expression for $x_{i+1}$ can be written as
 \begin{equation}
 x_{i+1} = x_i + \rho_i,
 \end{equation}
 and an expression for $x_n$ can be written as a sum of random numbers,
 \begin{equation}
 x_{n} = x_0 + \sum_{i=1}^{n} \rho_i,
 \end{equation}
 where $x_0$ is the initial position at time $t=0$.

 \subsection{Extending to two dimensions}

 To extend the stochastic differential equation from one to two dimensions, we can consider the solution to the unbounded two dimensional diffusion equation,
 \begin{equation}
 \phi(x,y,t) = \frac{1}{4\pi Dt}\exp \left(\frac{-(x^2+y^2)}{4Dt} \right) .
 \label{eqn:diffusion_solution2d}
 \end{equation}
 By separating this into $x$ and $y$ directions, we see that the solution to the two dimensional diffusion model is simply two one dimensional diffusion solutions multiplied together,
 \begin{equation}
 \phi(x,y,t) = \frac{1}{\sqrt{4\pi Dt}}\exp \left(\frac{-x^2}{4Dt} \right) \frac{1}{\sqrt{4\pi Dt}}\exp \left(\frac{-y^2}{4Dt} \right) .
 \end{equation}
 By treating the $x$ and $y$ directions separately, we use \eqn{eqn:SDE} to generate two stochastic differential equations for movement in each direction,
 \begin{align}
 dx_i &= \rho_i, \\
 dy_i &= \lambda_i,
 \end{align}
 where $\{\rho_i\}$ and $\{\lambda_i\}$ are both sets of random numbers chosen from the normal distribution with zero mean and standard deviation $\sqrt{2D\tau}$, such that $\rho_i \sim \mathcal{N}(0,2D\tau)$ and $\lambda_i \sim \mathcal{N}(0,2D\tau)$. Expressions for $x_n$ and $y_n$ can be written as sums of random numbers,
 \begin{align}
 x_{n} = x_0 + \sum_{i=1}^{n} \rho_i, \\
y_{n} = x_0 + \sum_{i=1}^{n} \lambda_i,
 \end{align}
 where $(x_0,y_0)$ is the initial position at time $t=0$.

\begin{figure} [h]
     \centering
         \includegraphics[width=0.8\textwidth]{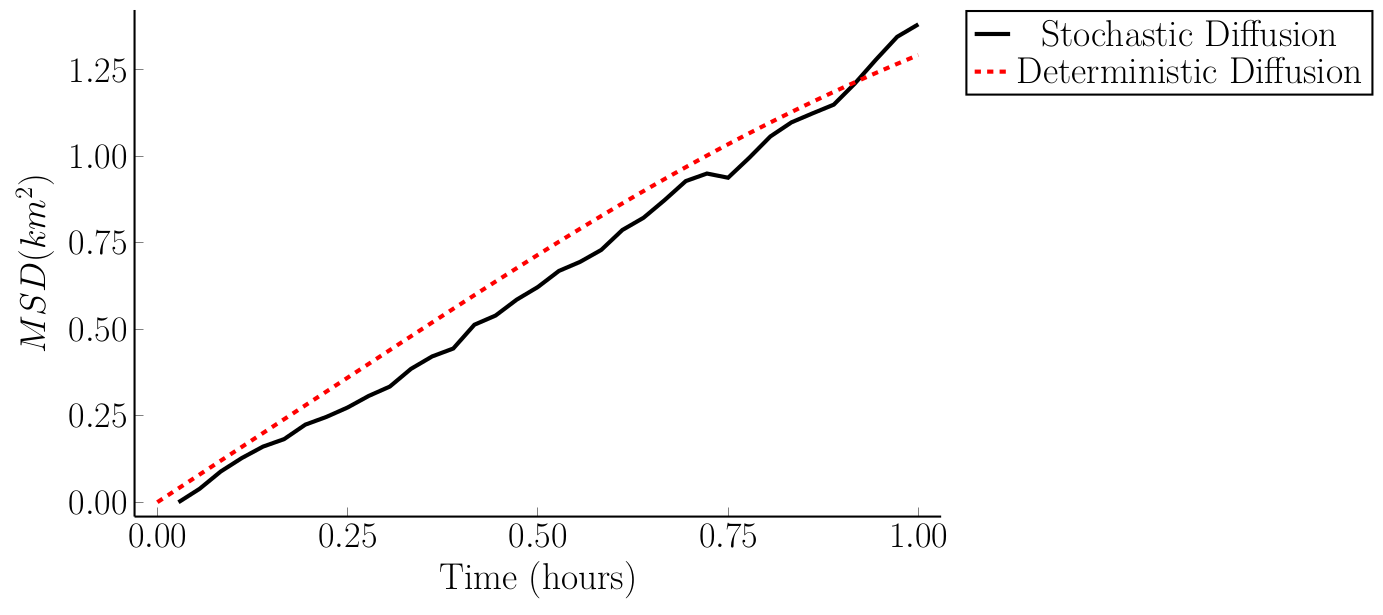}
         \caption{The mean squared displacement for stochastic and deterministic diffusion models in two dimensions with diffusion coefficient $D = 100\mathrm{m^2s^{-1}}$. The stochastic simulation uses $N = 100$ bats and a time step of $\tau$ = 10.}
     \label{fig:stochastic_deterministic_diffusion}
 \end{figure}

 The MSD for a stochastic diffusion simulation is shown in \fig{fig:stochastic_deterministic_diffusion}, along with the MSD for a deterministic diffusion simulation with the same parameters. The plot shows that both simulations give similar results, however the MSD for the stochastic model fluctuates around the value from the deterministic due to stochasticity.

 \subsection{A stochastic model for phase 2, return to the roost}
 As discussed in \sect{sect:convection}, the convection-diffusion does not provide a good model for phase 2 because the bats closest to the roost reach the roost first and stop moving. In this section we will derive a model that solves this problem by selecting only the bats furthest from the roost to drift back. All bats will undergo diffusion throughout the night, and during phase 2 the bat furthest from the roost will also undergo convection towards the roost.

\subsubsection{A stochastic convection-diffusion model}
 We will use a two dimensional convection-diffusion model in cartesian coordinates,
 \begin{multline}
   \D{\phi(x,y,t)}{t} = D \left(\DD{\phi(x,y,t)}{x} + \DD{\phi(x,y,t)}{y} \right) - \\ \left(\frac{x}{\sqrt{x^2+y^2}}\chi(x,y,t)\D{\phi(x,y,t)}{x} +\frac{y}{\sqrt{x^2+y^2}}\chi(x,y,t)\D{\phi(x,y,t)}{y} \right),
   \label{eqn:diffusion_cartesian2dstochastic}
 \end{multline}
where the expressions $x/\sqrt{x^2+y^2}$ and $y/\sqrt{x^2+y^2}$ ensure that convection acts towards the roost at $(x=0,y=0)$. As in \sect{sect:convection}, the expression for the convection coefficients is time-dependent, and also dependent on the distance from the roost,
 \begin{equation}
 \chi (x,y,t) =  \begin{cases}
     0, & \text{for } t < T, \\
     \sqrt{x^2+y^2}^{\beta}\chi_0, & \text{for } t \ge T.
   \end{cases}
   \label{eqn:stochastic_chi}
 \end{equation}
 The stochastic differential equations can be constructed using a superposition of diffusion and convection effects. The convection coefficient is equivalent to a drift velocity, convection causes a displacement in the $x$ and $y$ directions in one timestep $\tau$ of $x\chi(x,y,t)\tau/\sqrt{x^2+y^2}$ and $y\chi(x,y,t)\tau/\sqrt{x^2+y^2}$ respectively. The expressions for $x_{i+1}$ and $y_{i+1}$ can then be written as
 \begin{align}
 x_{i+1} = x_i + \rho_i - \frac{x_i}{\sqrt{x_i^2+y_i^2}}\chi(x_i,y_i,t)\tau, \\
 y_{i+1} = y_i + \lambda_i - \frac{y_i}{\sqrt{x_i^2+y_i^2}}\chi(x_i,y_i,t)\tau,
 \end{align}
 where ${\rho_i}$ and ${\lambda_i}$ are sets of random numbers such that $\rho_i \sim \mathcal{N}(0,2D\tau)$ and $\lambda_i \sim \mathcal{N}(0,2D\tau)$ as before.

 \subsubsection{A `leapfrog' model}
 We can now use this convection model to derive a model that selects only the bats furthest from the roost at each timestep to drift back towards the roost in order to solve the problem of the closest bats returning to the roost first. The convection coefficient is given by \eqn{eqn:stochastic_chi} for only the furthest bat at each timestep and is set to zero for all other bats. At each timestep, the convection coefficient is calculated as follows. First, find the index of bat at the furthest distance from roost,
 \begin{equation}
 j_{max} = \argmax_{j \in n} \left( \sqrt{x_j^2+y_j^2} \right).
 \end{equation}
 The convection coefficient is 0 for $t<T$, during phase 1, and for $t>T$ it is non-zero only for the furthest bat,
 \begin{equation}
 \chi_j (x,y,t) =  \begin{cases}
     0, & \text{for } t < T, \\
     \sqrt{x^2+y^2}^{\beta}\chi_0, & \text{for } t \ge T \text{ and } j = j_{max}.
   \end{cases}
   \label{eqn:r_i_dependentchi_stochastic}
 \end{equation}
As we cannot calculate the diffusion coefficient for phase 2 of movement, this is also time-dependent,
 \begin{equation}
 D(t) =  \begin{cases}
    D_1, & \text{for } t < T, \\
    D_2, & \text{for } t \ge T ,
  \end{cases}
  \label{eqn:t_dependentD}
 \end{equation}
 where $D_1$ and $D_2$ are the diffusion coefficients for phase 1 and phase 2 respectively, and are both positive constants.

The stochastic differential equation for bat $j$ at timestep $i$ is then written as
\begin{align}
x_{i+1,j} = x_{i,j} + \rho_{i,j} - \frac{x_{i,j}}{\sqrt{x_{i,j}^2+y_{i,j}^2}}\chi_j(x_{i,j},y_{i,j},t)\tau, \\
y_{i+1,j} = y_{i,j} + \lambda_{i,j} - \frac{y_{i,j}}{\sqrt{x_{i,j}^2+y_{i,j}^2}}\chi_j(x_{i,j},y_{i,j},t)\tau,
\end{align}
where ${\rho_{i,j}}$ and ${\lambda_{i,j}}$ are sets of random numbers such that $\rho_{i,j} \sim \mathcal{N}(0,2D(t)\tau)$ and $\lambda_{i,j} \sim \mathcal{N}(0,2D(t)\tau)$ as before.

\begin{figure} [h]
     \centering
         \includegraphics[width=0.8\textwidth]{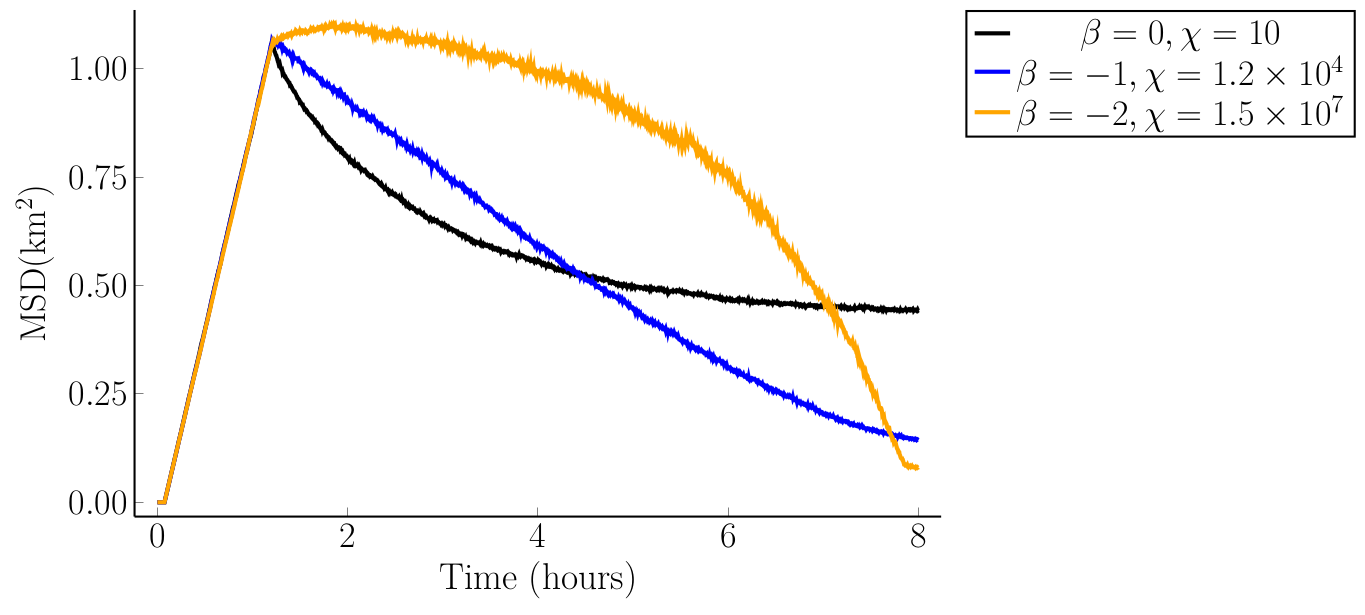}
         \caption{The mean-squared displacement for various exponents $\beta$. The parameters for these simulation were $D_1 = 65\mathrm{m^2s^{-1}}$, $D_2 = 46.5\mathrm{m^2s^{-1}}$ and $T=4350$ seconds.}
     \label{fig:leapfrog_beta}
 \end{figure}

The MSD for simulations with $n=100$ bats with various exponents $\beta$ are shown in \fig{fig:leapfrog_beta}. The curves for $\beta=0$ and $\beta = -1$ are convex, however for $\beta = -2$ the curve is concave, and a similar shape to the radio tracking data.

\begin{figure}
     \centering
     \begin{subfigure}[b]{0.47\textwidth}
         \centering
         \includegraphics[width=\textwidth]{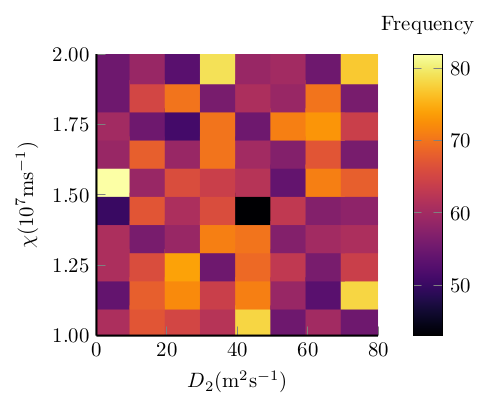}
         \caption{Prior joint distribution for $\chi$ and $D_2$.}
         \label{fig:stochastic_prior}
     \end{subfigure}
     \hfill
     \begin{subfigure}[b]{0.52\textwidth}
         \centering
         \includegraphics[width=\textwidth]{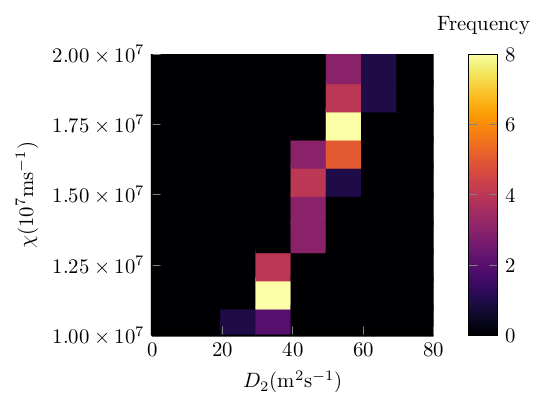}
         \caption{Posterior joint distribution for $\chi$ and $D_2$.}
         \label{fig:stochastic_posterior}
     \end{subfigure}
     \caption{2D histograms of prior and posterior joint distributions for convection and diffusion coefficients $\chi$ and $D_2$}
     \label{fig:stochastic_prior_post}
\end{figure}

To compare the model to the radio tracking data, the diffusion coefficient for phase 1 was calculated using
 \texttt{LsqFit.jl}, a package for least squares fitting in Julia \cite{LsqFit}. A
 straight line was fit to the initial linear segment, for time $0 \leq t < 3000$ seconds, and the gradient was used
 along with \eqn{eqn:diffusion_msd2d} to determine the diffusion coefficient as $D_1
 = 63.4 \mathrm{m^2s^{-1}}$. For phase 2, the exponent $\beta = -2$ was used, and the parameters $D_2$ and $\chi$ were fit using ABC. In this case, $\bm{Y}$ is the MSD at each point in time, and $\bm{X}$ is the expected MSD at each time
 point for parameters $\theta' = (D_2',\chi') $, calculated using the model for
 diffusion on a shrinking domain. The distance metric $\rho(\bm{X},\bm{Y})$ is the coefficient of determination,
 \begin{equation}
 r^2 = 1 - \frac{\sum_i(y_i - x_i)^2 }{\sum_i (y_i - \overline{y})^2},
 \end{equation}
 where $y_i$ corresponds to each value in $\bm{Y}$ and $x_i$ corresponds to each value in $\bm{X}$. The ABC algorithm was run for a sample size of $n = 5000$, and $\epsilon$ was chosen such that the best 1\% of parameter values were added to the posterior. The prior and posterior distributions are shown in \fig{fig:stochastic_prior_post}. The posterior distribution shows that the parameters $D_2$ and $\chi$ are clearly correlated. The estimate for each parameter is calculated by taking the mean of the posterior, $D_2 = 46.5\mathrm{m^2s^{-1}}$ seconds and $\chi = 1.5 \times 10^7\mathrm{ms^{-1}}$.

\begin{figure} [h]
    \centering
        \includegraphics[width=0.8\textwidth]{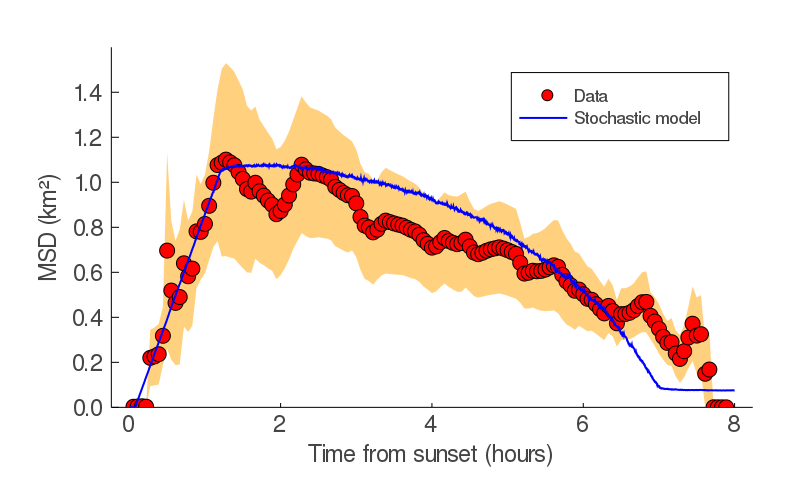}
        \caption{MSD for the stochastic model compared to radio tracking data. For the stochastic model, the MSD shown is the median value at each timestep over 1000 simulations.}
    \label{fig:leapfrog_fit}
\end{figure}

\begin{figure} [h]
    \centering
        \includegraphics[width=0.7\textwidth]{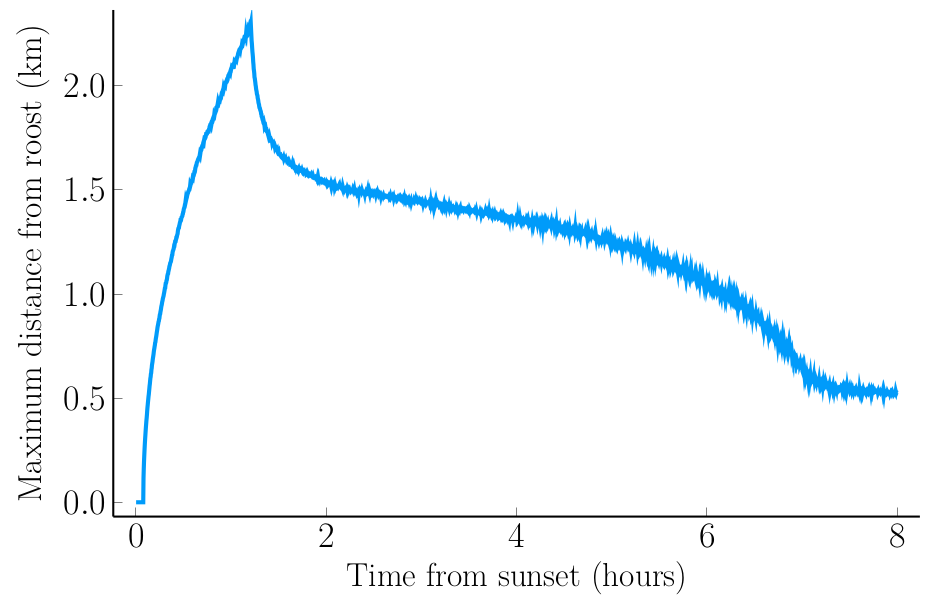}
        \caption{The maximum distance from the roost over time for the simulation shown in \fig{fig:leapfrog_fit}.}
    \label{fig:stochastic_maxd}
\end{figure}

 The results of a simulation with the parameters calculated with ABC is shown in \fig{fig:leapfrog_fit} to compare with the radio tracking data. The curve provides a good fit to the radio tracking data, suggesting that this model provides a good description of bat behaviour when foraging. The maximum distance from the roost over time for the same simulation is plotted in \fig{fig:stochastic_maxd}, showing that during phase 2 of movement the maximum distance from the roost is decreasing. As diffusion is unbounded for this simulation, some bats travelled much further from the roost than the rest of the colony during phase 1, and the initial steep decrease in phase 2 is due to these outliers drifting back quickly.

 Next we will use this leapfrog model to inform a deterministic model for phase 2, using diffusion on a shrinking domain, in which the maximum possible distance from the roost decreases over time.

\subsubsection{Diffusion on a shrinking domain} \label{shrink}
 Next we will consider diffusion on a shrinking domain, assuming that bats disperse at the beginning of the night, but tend to move back towards the roost, narrowing the area in which they forage as the night goes on. To model diffusion on a shrinking domain, we can consider a frame of reference that moves with a flow caused by the domain shrinking \cite{crampin1999reaction}. Considering an elemental volume $w(t)$, the velocity field $\bm{a}$ of the flow at position $\bm{X}$ is
\begin{equation}
\bm{a}(\bm{X},t) = \frac{d\bm{X}}{dt}.
\end{equation}
\begin{figure} [h]
    \centering
        \includegraphics[width=0.5\textwidth]{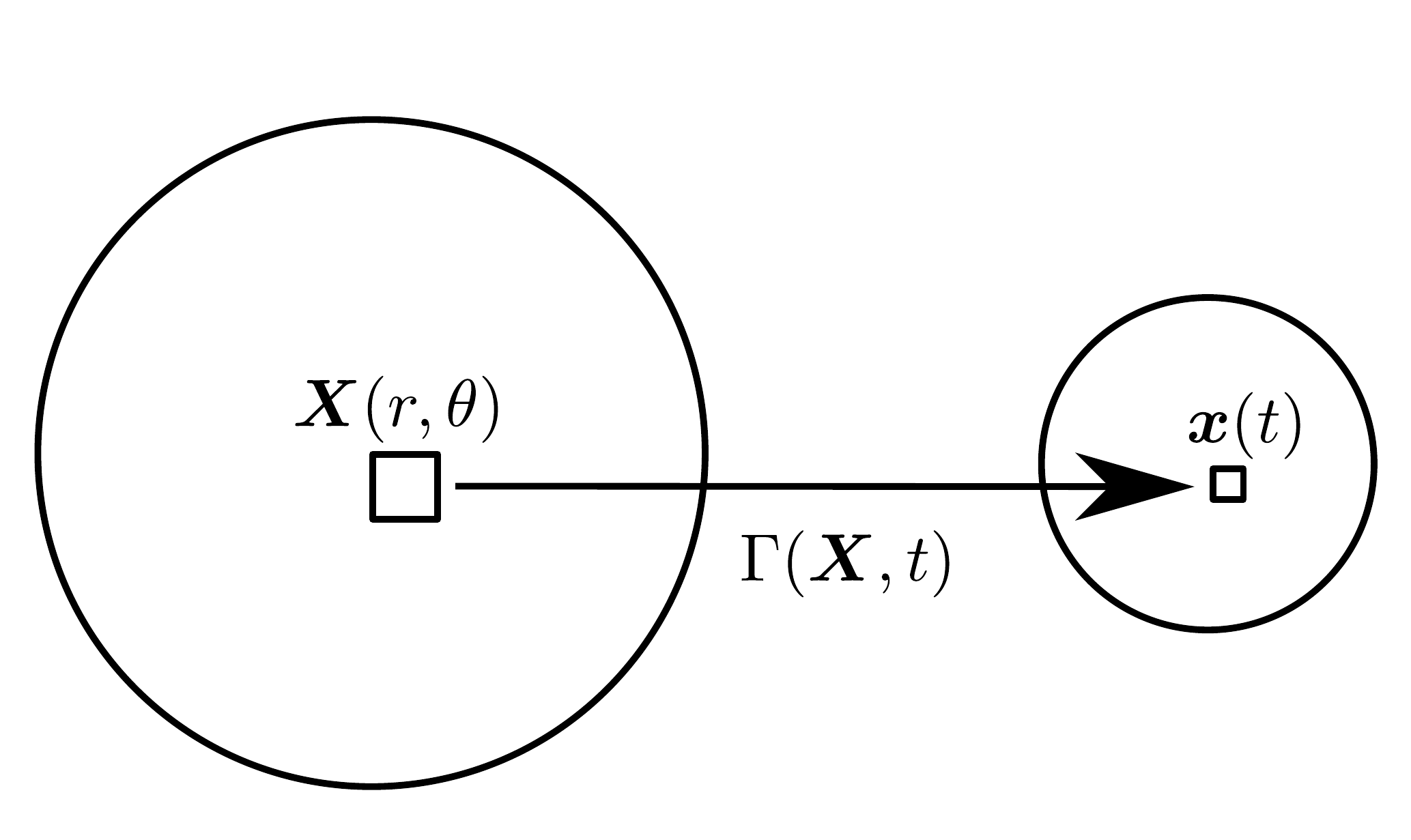}
        \caption{A diagram showing the mapping of the stationary to the shrinking domain frame of reference. }
    \label{fig:shrink_diagram}
\end{figure}
Moving from the stationary to the shrinking domain frame of reference requires a Lagrangian description of the domain which maps each point in the domain from the stationary frame to the shrinking frame. If $\bm{X} = (r(t=0),\theta(t=0))$ is the initial location of an element $w(t)$ and $\bm{x} = (r(t),\theta(t))$ is the location of the element at time $t$ then the mapping is defined by the function $\Gamma(\bm{X},t) $ as $\bm{x}(t) =\Gamma(\bm{X},t) $. A diagram showing the mapping is displayed in \fig{fig:shrink_diagram}. For a growth rate $l(t)$, the mapping function is
\begin{equation}
\Gamma(X,t) = Xl(t).
\label{eqn:mapping}
\end{equation}
Then, the velocity field is defined by
\begin{equation}
\bm{a}(\bm{X},t) = \frac{d\bm{\Gamma}(\bm{X},t)}{dt} .
\label{dgammadt}
\end{equation}
Using the chain rule to expand \eqn{dgammadt} gives
\begin{equation}
    \Dd{\Gamma_i}{t}{X_k} = \sum_{j=1}^{3}\D{a_i}{x_j}\D{\Gamma_j}{X_k},
\end{equation}
where $\Gamma_i$ and $a_i$ are the $i^{th}$ components of $\bm{\Gamma}$ and $\bm{a}$ \cite{crampinnonuniform}.
Considering the mapping function in \eqn{eqn:mapping}, a stationary element is mapped onto the shrinking domain with $\bm{x}(t) = \bm{X}l(t)$. The diffusion equation in polar coordinates, \eqn{eqn:diffusion_polar2d}, can then be transformed from the shrinking domain variables by mapping the derivatives to the new domain using
\begin{equation}
\at{\D{}{t}}{x} = \at{\D{}{t}}{X} - X \frac{\dot{l}(t)}{l(t)} \D{}{X}
\end{equation}
and
\begin{equation}
\D{}{x} = \frac{1}{l(t)}\D{}{X}.
\end{equation}
The scaled diffusion equation in the stationary frame of reference is then given by
\begin{equation}
\D{\phi}{t} = \frac{D}{Xl(t)^2}\D{}{X}\left(X \D{\phi}{X} \right) + X \frac{\dot{l(t)}}{l(t)}\D{\phi}{X}.
\end{equation}
\begin{figure} [h]
    \centering
        \includegraphics[width=0.3\textwidth]{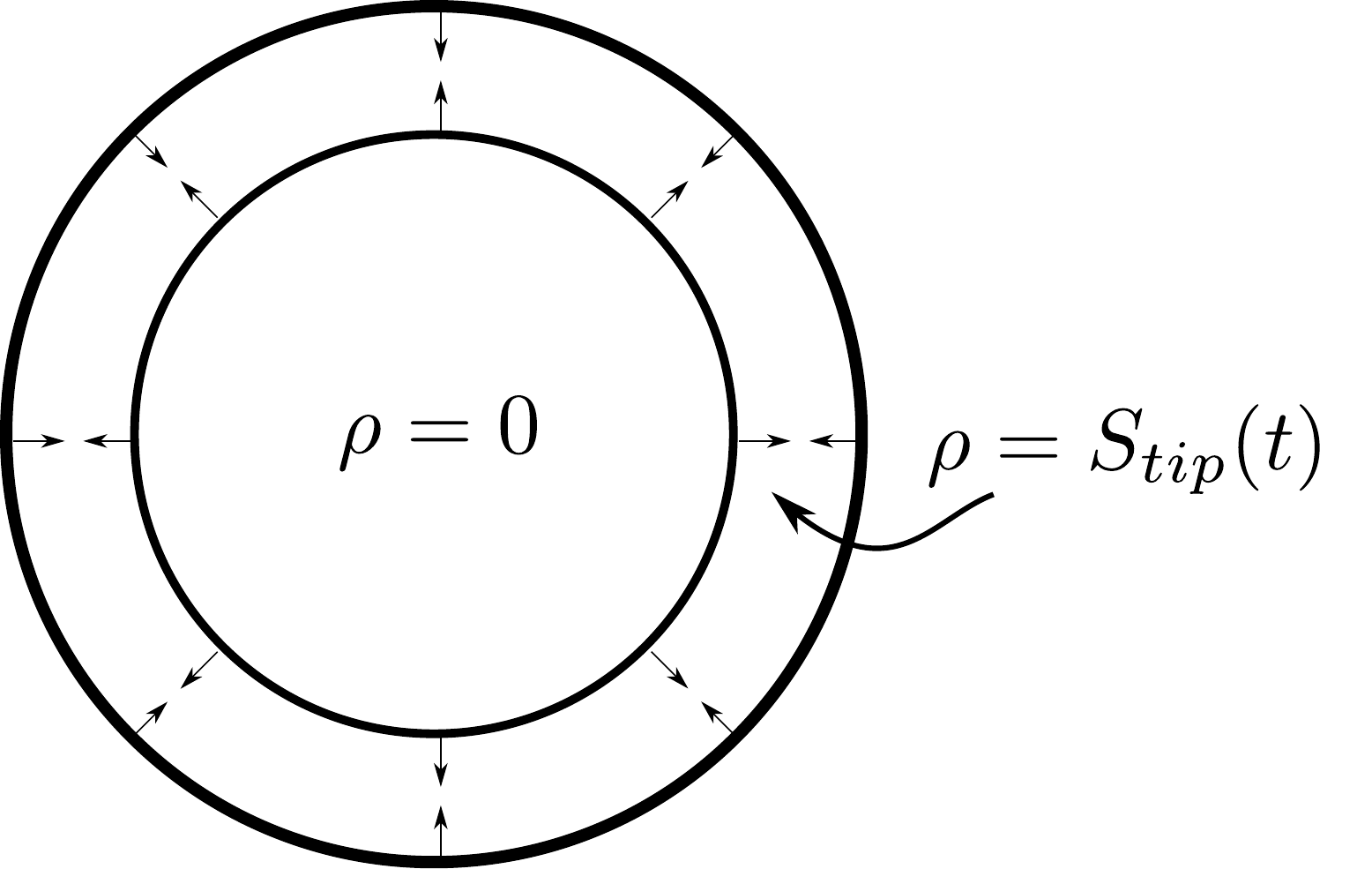}
        \caption{A diagram showing a domain shrinking apically, such that shrinking is restricted to the edge of the domain.}
    \label{fig:apical_diagram}
\end{figure}
We will consider a domain shrinking apically, in which shrinking is restricted to a region of width $\delta$ at the tip of domain. The shrinking rate $\rho$ is zero everywhere except at the edge of the domain,
\begin{equation}
\D{a}{r} = \rho = \begin{cases}
		0, & 0 \leq r \leq R(t) - \delta, \\
		S_{tip}(t), & R(t) - \delta \leq r \leq R(t), \\
		\end{cases}
        \label{eqn:s_tip}
\end{equation}
where $S_{tip}(t)$ is the shrinking rate in the element at the edge of the boundary. A diagram illustrating \eqn{eqn:s_tip} is shown in \fig{fig:apical_diagram}. Therefore, the size of the domain $R(t)$ is given by $R(t) = 1 + \delta \int_0^t S_{tip}(t')dt'$. Considering the shrinking region to be much smaller than the domain size, the system can be reduced to an Eulerian moving boundary problem.
Assuming also that the growth rate $l$ is small, the equation reduces to the diffusion equation
\begin{equation}
\D{\phi}{t} = \frac{D}{Xl(t)^2}\D{}{X}\left(X \D{\phi}{X} \right).
\end{equation}
If the diffusion rate is larger than the rate at which
 the domain changes size, the solution should approximate steady state diffusion. As shown in \sect{phase1}, the solution to the diffusion equation on a bounded domain tends to uniformity. On a circular domain, as $t\rightarrow \infty$,
  \begin{equation}
  \phi \rightarrow \frac{1}{\pi R^2}.
  \label{eqn:uniform_circle}
  \end{equation}
Thus, we expect the probability distribution to remain
 approximately uniform over the domain and the expected MSD at time $t$ can be calculated using this probability distribution,
 \begin{align}
 \left<r^2\right> 	&= \int_{\Omega}r^2 \phi(r,t) d\Omega , \nonumber\\
                 	&= \int_0^{2\pi}\int_0^{R(t)} \frac{r^3}{\pi R(t)^2} dr d\theta, \nonumber \\
	                &= \frac{R(t)^2}{2} .
 \label{eqn:shrink_domain}
 \end{align}
   \begin{figure} [h]
       \centering
           \includegraphics[width=0.3\textwidth]{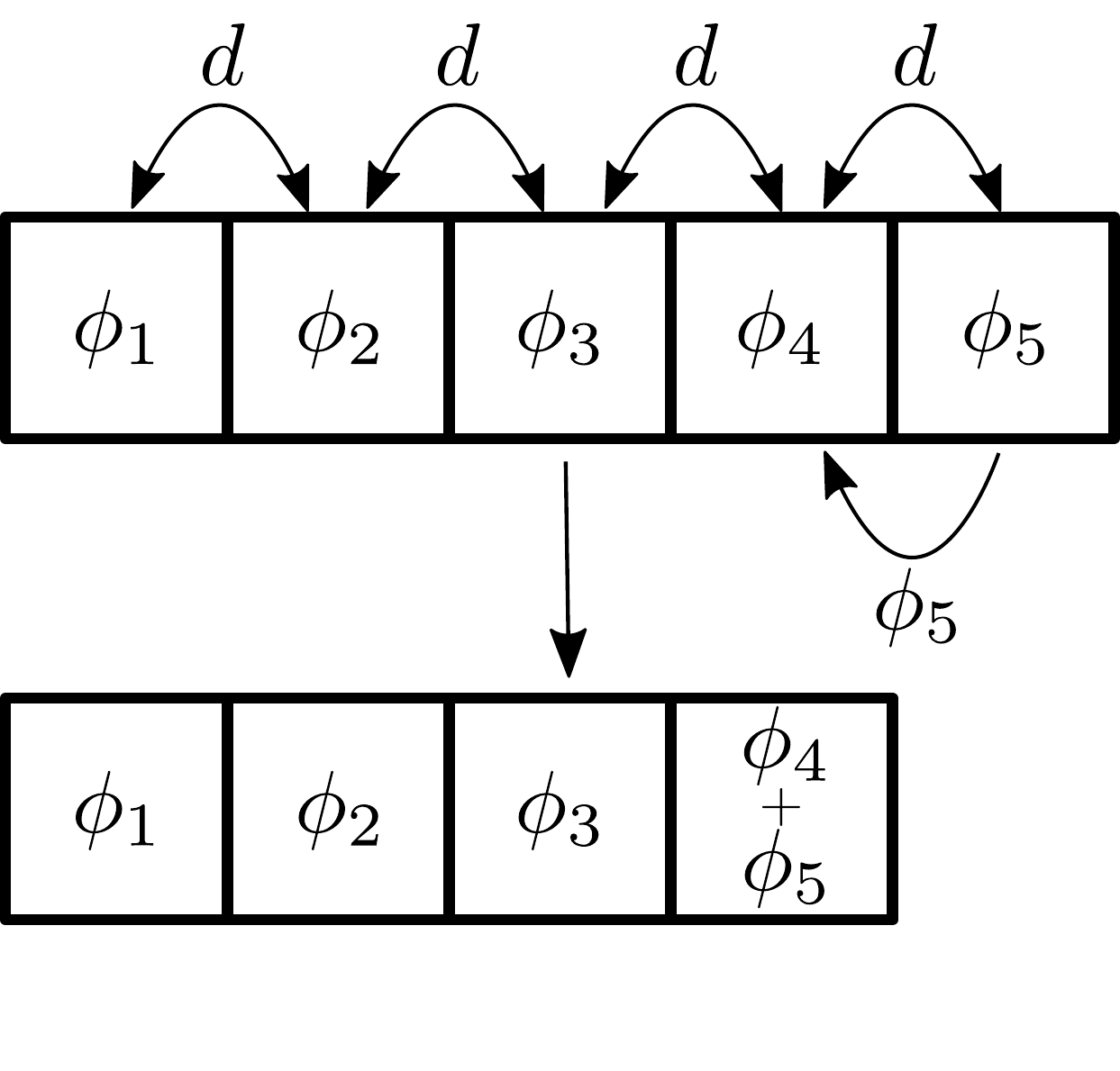}
           \caption{A diagram illustrating a discretised diffusion process on a shrinking domain. The diffusion process is denoted by $d$ and the probability density in annulus $i$ is denoted by $\phi_i$}
       \label{fig:shrinking_diagram}
   \end{figure}
\subsubsection{A discretised diffusion simulation on a shrinking domain}
%NOTE timescale?
The diffusion process on a shrinking domain can be simulated using the diffusion model described in \sect{phase1}. The simulation consists of two separate stages at each time step, illustrated in \fig{fig:shrinking_diagram}. First, the diffusion process moves probability density between annuli. Then, the size of the domain is reduced by adding the concentration in annulus $N$ to annulus $N-1$ and removing annulus $N$.

Since the MSD for this phase is a negative parabola, the shrinking rate is chosen to give
 \begin{equation}
 \left<r^2\right> \propto a - t^2.
 \label{eqn:MSD_shrink}
 \end{equation}
From \eqn{eqn:shrink_domain}, a time dependent domain size $R(t)$ of
\begin{equation}
R(t) = \sqrt{2(R_0^2 - \alpha t^2)}
\label{eqn:Rt}
\end{equation}
gives an expected MSD of
 \begin{equation}
 \left<r^2\right> = {R_0}^2 -\alpha t^2.
 \end{equation}

\begin{figure} [h]
        \centering
            \includegraphics[width=0.6\textwidth]{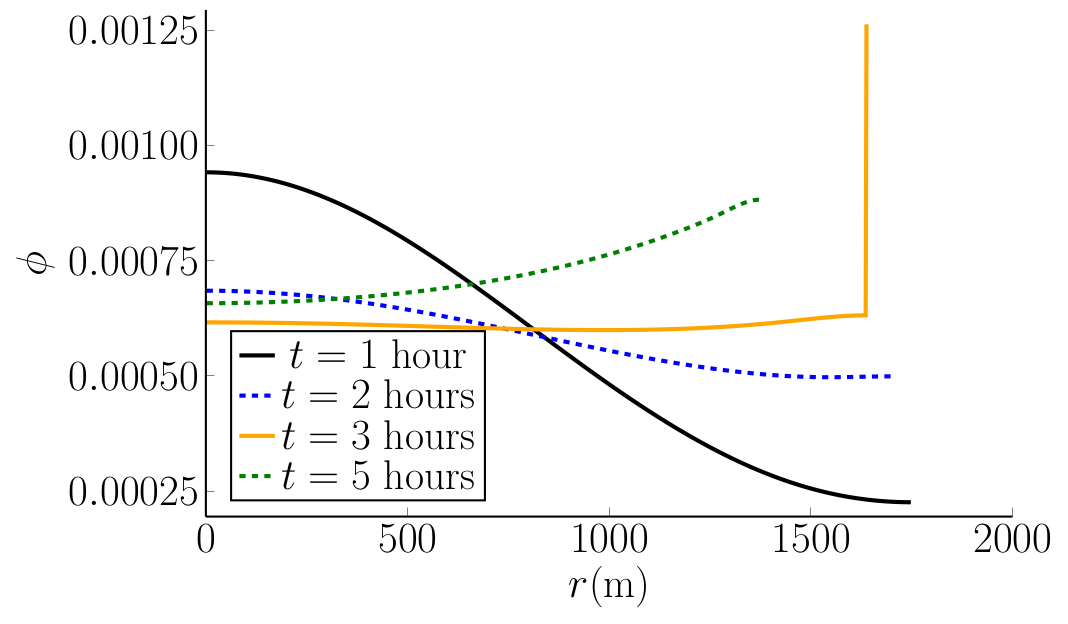}
            \caption{The probability density $\phi$ for diffusion simulation on a shrinking domain at $t = 1$, 4 and 6 hours. The initial condition is a delta function at $r = 0$ and the
             parameters for this simulation are $N = 100$, $D = 100\mathrm{m^2s^{-1}}$, $R_0 = 1800$m, $t_s = 1000$ seconds and $\alpha  = 3.24 \mathrm{m^2s^{-2}}$.}
        \label{fig:shrink_phi}
    \end{figure}

The result of a simulation of diffusion on a domain shrinking with rate $R(t)$ given by \eqn{eqn:Rt} is shown in \fig{fig:shrink_phi}. The probability density spreads across the domain due to the diffusion process, however the shrinking causes the probability density to increase at the right edge of the domain. In this case, the rate of diffusion is slower than the rate at which the domain shrinks, and therefore the probability density does not spread evenly across the domain.
%NOTE Add another plot to show both extremes: slow shrink rate, high diffusion. high shrink rate, slow diffusion.

\section{Comparison of convection-diffusion and shrinking domain models}

\begin{figure} [h]
    \centering
        \includegraphics[width=0.8\textwidth]{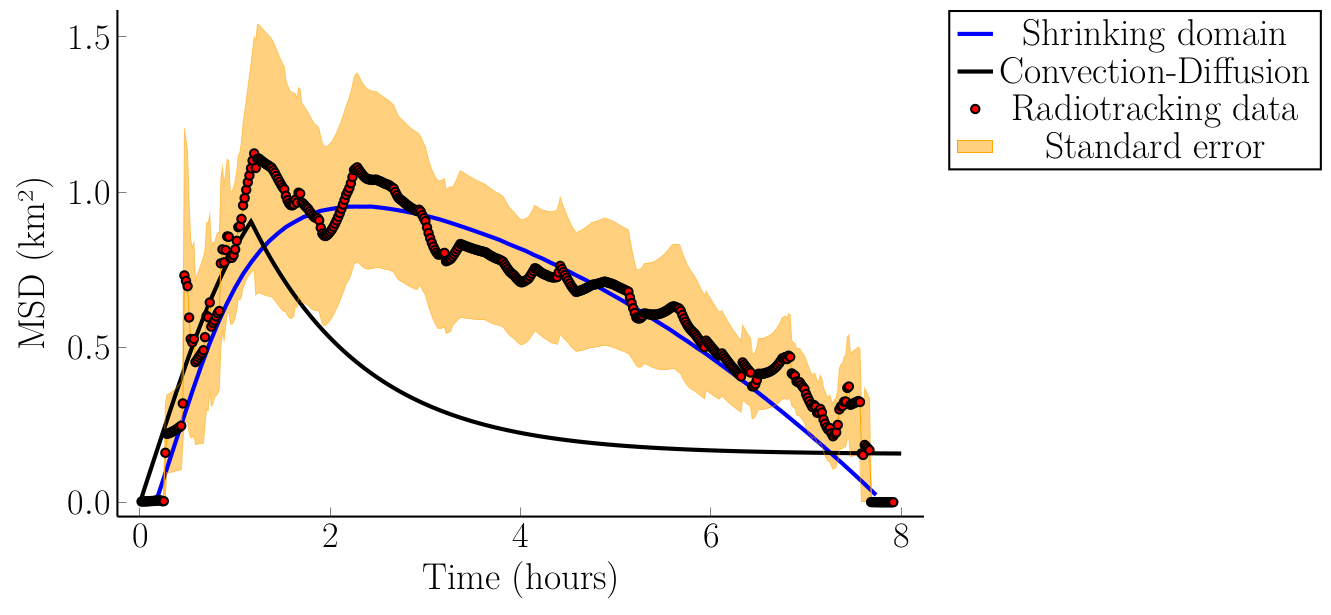}
        \caption{Comparison of the mean squared distance for a discrete convection-diffusion model and diffusion on a shrinking domain.
        }
    \label{fig:c-d_shrink}
\end{figure}

In order to compare the shrinking domain and convection-diffusion models, simulations were run for each model. For the convection-diffusion model, the diffusion process is first simulated for $T = 4000$ seconds, then the convection-diffusion process is simulated until the end of the night. The parameters for this simulation are $R = 2000$m, $N = 100$,
$D = 65\mathrm{m^2s^{-1}}$, and $\chi =  - 0.15\mathrm{ms^{-1}}$. For the shrinking domain model, the parameters used were $N = 100$, $D = 65\mathrm{m^2s^{-1}}$, $R_0 = 1800\mathrm{m}$, $t_s = 1000$ seconds and $\alpha  = 3.24 \mathrm{m^2s^{-2}}$. The mean-squared distance was calculated numerically, using a trapezium rule
approximation for the expectation value in \eqn{eqn:MSD_expectation}, and the results are shown in \fig{fig:c-d_shrink}.

\begin{figure}
     \centering
     \begin{subfigure}[b]{0.48\textwidth}
         \centering
         \includegraphics[width=\textwidth]{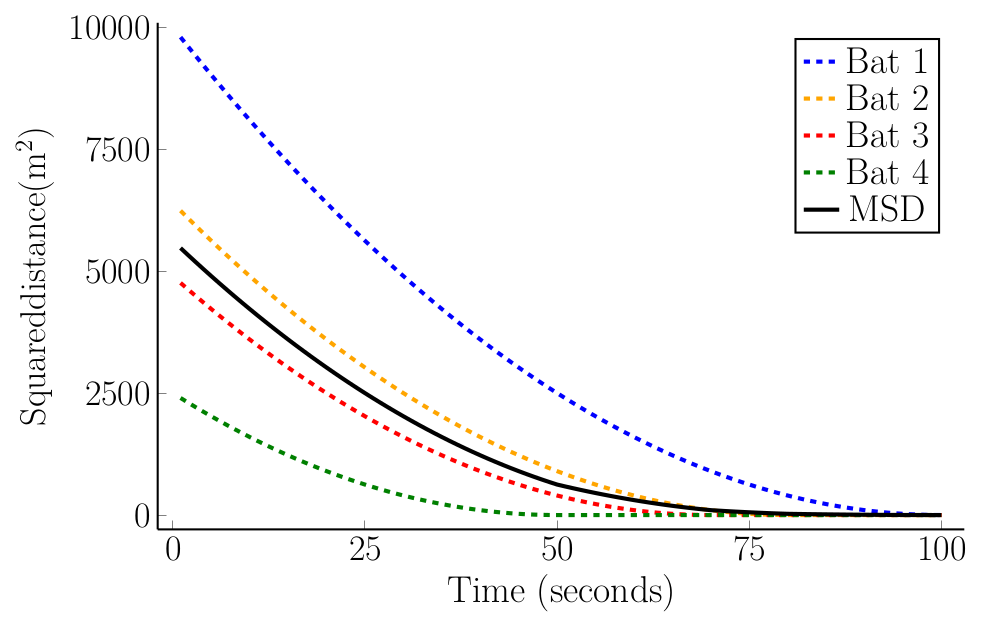}
         \caption{Linear convection with constant convection coefficient $\chi=1\mathrm{ms^{-1}}$.}
         \label{fig:convectionlinearchi}
     \end{subfigure}
     \hfill
     \begin{subfigure}[b]{0.48\textwidth}
         \centering
         \includegraphics[width=\textwidth]{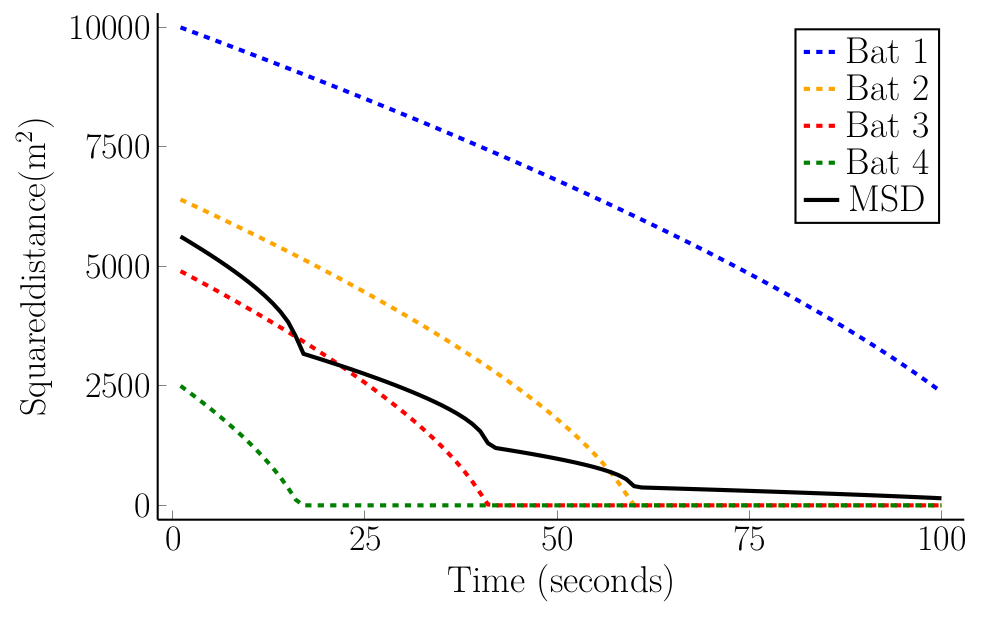}
         \caption{Distance dependent convection, $\chi_0= 3000\mathrm{ms^{-1}}$ and $\beta = -2$. }
         \label{fig:convectionbetaphi}
     \end{subfigure}
     \caption{The squared distance for 4 bats undergoing convection towards a roost with differing start points. Bat 1 starts at $r=100$m, bat 2 starts at $r=80$m, bat 3 starts at $r=70$m and bat 4 starts at $r=50$m.}
     \label{fig:convection_diagrams}
     \end{figure}

The result for the convection-diffusion model shows an initial linear dispersal, due to diffusion, followed by a sharp decrease as the convection process begins. The decrease slows and the curve flattens over time as probability density gathers at the boundary at $r=0$. Plots showing the squared distance for 4 bats undergoing convection are shown in \fig{fig:convection_diagrams}. For linear convection, the squared distance for each bat is concave, but for distance dependent convection with $\chi \propto r^{-2}$, the squared distance for each bat is convex. However, the bats closest to the roost return first and stop moving, and when we take the mean squared distance it produces a concave curve. When bats are also subject to diffusion, the mean squared displacement never reaches zero because diffusion continues to spread probability density across the domain and diffusion and convection are balanced. The shape of the convection curve is clearly inconsistent with the convex shape that arises from the radio tracking data in \fig{fig:MSD}, and convection-diffusion does not provide a good model for the radio tracking data. The shrinking domain model gives a very different result, an initial straight line dispersal which then begins to level off and decrease slowly, a similar shape to the radio tracking data.

\section{Model validation using radio tracking data}

The diffusion on a shrinking domain model was fit to the radio tracking data. The diffusion coefficient for the initial, linear dispersal was calculated using
\texttt{LsqFit.jl}, a package for least squares fitting in Julia \cite{LsqFit}. A
straight line was fit to the initial linear segment, for time $0 \leq t < 3000$ seconds, and the gradient was used
along with \eqn{eqn:diffusion_msd2d} to determine the diffusion coefficient as $D
= 63.4 m^2s^{-1}$. For the return phase, the shrinking rate was chosen to give a negative parabola for the MSD, as in \eqn{eqn:Rt}. The parameters to fit are $\alpha$, the rate at which the domain shrinks,
$t_s$, the time at which the domain begins to shrink and $R_0$, the initial size
of the domain. Since bats return to the roost at sunrise, the MSD is 0 at the end of the night. Therefore, $\alpha$ was chosen from $R_0$ to ensure the domain size shrinks to 0 at sunrise,
\begin{equation}
\alpha = \frac{R_0^2}{T^2},
\end{equation}
where $T = 8 $ hours is the time to sunrise.

 The parameters $R_0$ and $t_s$ were fit using Approximate Bayesian Computation (ABC). In this case, $\bm{Y}$ is the MSD at each point in time, and $\bm{X}$ is the expected MSD at each time
point for parameters $\theta' = (R_0',t_s') $, calculated using the model for
diffusion on a shrinking domain. The distance metric $\rho(\bm{X},\bm{Y})$ is the coefficient of determination,
\begin{equation}
r^2 = 1 - \frac{\sum_i(y_i - x_i)^2 }{\sum_i (y_i - \overline{y})^2},
\end{equation}
where $y_i$ corresponds to each value in $\bm{Y}$ and $x_i$ corresponds to each value in $\bm{X}$.

As there is initially no information about parameters $\theta$, the prior distribution for each parameter is assumed to be uniform over plausible values. For $t_s$, $p(t_s) \sim \unif(0,5000)$ seconds, and for the shrinking rate $R_0$, $p(R_0) \sim \unif(1500\mathrm{m},2500\mathrm{m})$. The posterior distribution $ p(\theta \mid \bm{Y})$ will be a distribution describing the probability of each set of possible parameters $\theta$, and is given by the mean value for each parameter $\overline{\bm{\theta}}$, where $\bm{\theta_i}$ is the $i$-th accepted sample.

\begin{figure} [h]
    \centering
        \includegraphics[width=\textwidth]{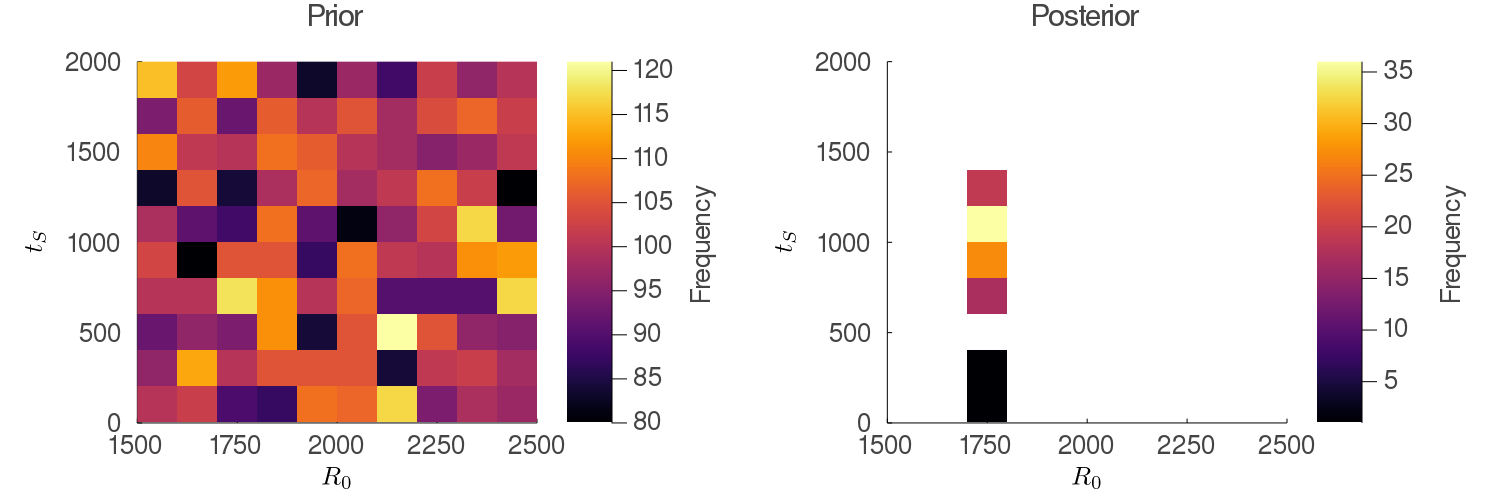}
        \caption{2D histograms of prior and posterior joint distributions for $R_0$ and $t_s$.}
    \label{fig:posterior}
\end{figure}

The ABC algorithm was run for a sample size of $n = 10^4$, and $\epsilon$ was chosen such that the best 1\% of parameter values were added to the posterior. The prior and posterior distributions are shown in \fig{fig:posterior}. The posterior distribution shows a very narrow distribution in $R_0$, suggesting that the MSD is very sensitive to domain radius. The estimate for each parameter is calculated by taking the mean of the posterior, $t_s = 901$ seconds and $R_0 = 1756$m.

\begin{figure} [h]
    \centering
        \includegraphics[width=0.6\textwidth]{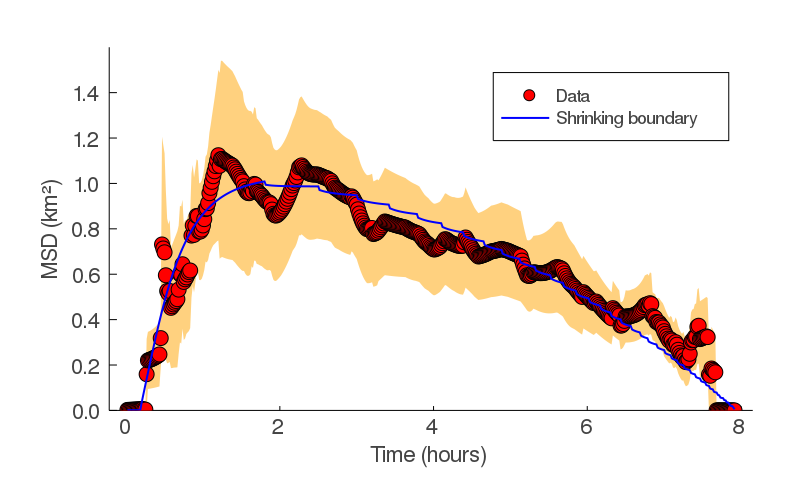}
        \caption{The MSD for a deterministic diffusion model on a shrinking domain of size $R(t) = R_0 - \sqrt{\alpha t^2}$. }
    \label{fig:fit}
\end{figure}

 The MSD simulation using fitted estimates of the true parameters is shown in \fig{fig:fit}. The curve shows a good fit with the radio tracking data, within the standard error for the majority of the night. The coefficient of determination calculated was $r^2 = 0.929$, suggesting that the model provides a good fit. Additionally, the distance from the roost for all recordings from the radio tracking data was calculated, and 93\% of recordings were within $R_0 = 1756$m, suggesting that this is a good estimate for maximum foraging radius for the majority of bats in this study.

\section{Discussion}
We have developed multiple mathematical tools that depend on Bayes theory and Approximate Bayesian Computation that allow us to model bat movement, parameterise these models and subsequently identify roost locations. Due to bats being important as an ecological health indicator and their protected status it is important to be able to find them so that their health can be studied and ensured.

Critically, we have presented a number of different models for bat motion dependent of multiple forms of reaction-diffusion equations. Specifically, we demonstrated that random motion is consistent to the movement during the first phase of the night's feeding, whilst leap-frogging, or domain shrinkage, is consistent with the motion in the second phase. We now discuss the interpretation of these motion strategies in terms of the bat's ecology.

The initial rapid dispersal from the roost in phase 1 can be explained by competition for resources. Bats who travel far from the roost have a larger area to themselves, and therefore more resources available to them. The model for phase 2 is not as easy to interpret, and there are a number of possible explanations.

\begin{figure} [h!!!t!!!b!!!p]
    \centering
        \includegraphics[width=0.6\textwidth]{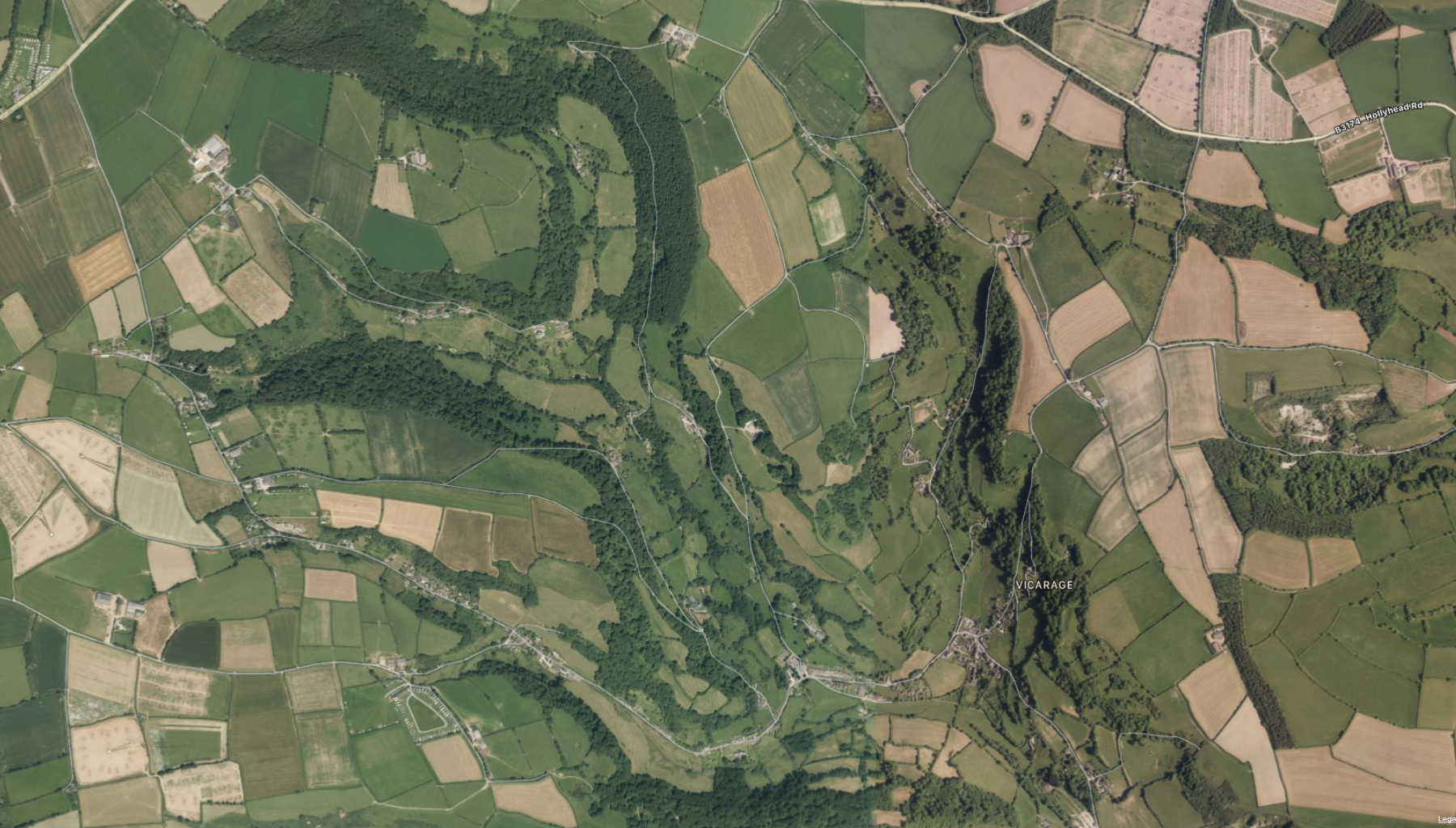}
        \caption{A satellite image of the area covered by bats in the survey. }
    \label{fig:satellite}
\end{figure}

The movement of bats within a domain with a shrinking boundary corresponds to those furthest from the roost always moving towards the roost, whilst those closer to the roost move diffusively. The probability distributions for this simulation in \fig{fig:shrink_phi} show a clustering of probability density against the moving boundary, suggesting that bats behaving in this way would gather together. This behaviour could be explained by a desire for bats to move towards the locations of other foraging bats. In fact, a common foraging strategy amongst some bat species is to eavesdrop on the hunting calls of other bats to easily and quickly locate hunting grounds \cite{roelekelandscape, egert2018resource}. This strategy is most common in landscapes dominated by cropland, where prey is difficult to find for a single bat due to patchy and ephemeral or unpredictable insect distribution and is uncommon in woodland where insect distribution is more reliable. Eavesdropping allows bats to locate areas with insects by following bats that have already found these hunting grounds. A satellite image of the area covered by bats in the survey is displayed in \fig{fig:satellite}, showing that the area is primarily farmland, with very little woodland, and thus it is likely that bats foraging in this landscape may employ this eavesdropping technique. If bats are using eavesdropping whilst foraging, we would expect to see bats gathering together over the course of the night. It is not possible to test this hypothesis using standard radio tracking techniques: we might record one tagged bat in a specific location, but there may be others around.

Additionally, bats may have a mind-map of where they want to forage (for example, routes learnt from their mothers), and decide to turn back towards the roost once they have reached the edge of their foraging zone. It would be possible to test this hypothesis by tracking one bat over multiple days to see if they repeat their foraging routes, however this is beyond the scope of the work discussed here. An alternative explanation is the possibility that bats do not want to be further from the roost than the rest of the colony the furthest from the roost once they have eaten, and therefore start to travel back after foraging.

\end{document}